\documentclass[a4paper,fleqn]{cas-sc}
\usepackage{hyperref}

\usepackage[numbers]{natbib}

\def\tsc#1{\csdef{#1}{\textsc{\lowercase{#1}}\xspace}}
\tsc{WGM}
\tsc{QE}
\tsc{EP}
\tsc{PMS}
\tsc{BEC}
\tsc{DE}

\begin{document}
\let\WriteBookmarks\relax
\def\floatpagepagefraction{1}
\def\textpagefraction{.001}

\newcommand{\eirik}[1]{\textcolor{red}{Eirik: ``#1''}}
\newcommand{\markl}[1]{\textcolor{black}{#1}}

\shorttitle{WAVEx}    

\shortauthors{Loveland \emph{et al.}}  

\title [mode = title]{WAVEx: Stabilized Finite Elements for Spectral Wind Wave Models Using FEniCSx}  



%

\author[1]{Mark Loveland}[orcid=0000-0002-2164-2884]

\cormark[1]


\ead{markloveland@utexas.edu}

\ead[url]{https://www.researchgate.net/profile/Mark-Loveland-2}

\credit{Methodology, Software, Validation, Writing - Original Draft}

\affiliation[1]{organization={Oden Institute for Computational Engineering and Sciences, University of Texas at Austin},
            addressline={201 E. 24th St.}, 
            city={Austin},
            state={Texas},
            postcode={78712}, 
            country={USA}}

\author[1,2,3]{Eirik Valseth}[orcid=0000-0001-6940-4191]



\credit{Methodology, Writing - Review \& Editing}

\affiliation[2]{organization={The Department of Data Science, The Norwegian University of Life Science},
            addressline={Drøbakveien 31}, 
            city={Ås},
            postcode={1433}, 
            country={Norway}}
\affiliation[3]{organization={Department of Scientific Computing and Numerical Analysis, Simula Research Laboratory},
            addressline={Kristian Augusts gate 23}, 
            city={Oslo},
            postcode={0164}, 
            country={Norway}}    
\author[4]{Jessica Meixner}
\credit{Methodology, Writing - Review \& Editing}
\affiliation[4]{organization={National Oceanic and Atmospheric Administration (NOAA), NWS, NCEP, EMC},
            addressline={830 University Research Court}, 
            city={College Park},
            state={MD},
            postcode={20740}, 
            country={USA}}

\author[1]{Clint Dawson}[orcid=0000-0001-7273-0684]
\credit{Resources, Supervision, Project administration}
\cortext[1]{Corresponding author}

\begin{abstract}
The \markl{prediction of the wind wave spectrum of the ocean using numerical models are an important tool for researchers, engineers, and communities living in coastal areas. The governing equation of the wind wave models, the Wave Action Balance Equation, presents unique challenges for implementing reliable numerical models because it is highly advective, highly nonlinear and high dimensional. Historically, most operational models have utilized finite difference methods, others have used finite volume methods but relatively few attempts at using finite element methods.} In this work, \markl{we seek to fill this gap by investigating} several \markl{different finite element} discretizations of the Wave Action Balance Equation. The methods, which include streamline upwind Petrov-Galerkin (SUPG), least squares, and discontinuous Galerkin, are implemented and convergence properties are examined for some simplified 2-D test cases. 
Then, a new spectral wind wave model, WAVEx, is formulated and implemented \markl{for the full problem setting. }
\markl{WAVEx  uses continuous finite elements along with SUPG stabilization in geographic/spectral space that allows for fully unstructured triangular meshes in both geographic and spectral space}. For propagation in time, a \markl{second order fully} implicit finite difference method is used. When source terms are active, a second order operator splitting scheme is used \markl{to linearize the problem}. In the splitting scheme, propagation is solved using the implicit method and the nonlinear source terms are treated explicitly. Several test cases, including analytic tests and laboratory experiments, are demonstrated and results are compared to analytic solutions, observations, as well as output from another model that is used operationally.
\end{abstract}


\begin{highlights}
\item Several finite element methods are investigated for solving the Wave Action Balance Equation (WAE) on simplified 2D problems
\item A finite element solver using SUPG stabilization, implicit time stepping, and Strang splitting for the full 4D WAE is implemented using FEniCSx
\item A set of six test cases from the Office of Naval  Research test bed are run using FEniCSx on unstructured meshes and accuracy is compared with analytic solutions and SWAN output
\end{highlights}


\maketitle

\section{Introduction}\label{sec:intro}

 Wind waves are those that typically range in period from 0.25-30 seconds, or wave lengths of 0.1-1500 meters, and are the product of disturbances of the ocean surface due to locally generated winds~\cite{holthuijsen_2007}. Wind wave models take into consideration the local winds, bathymetry, and currents in order to approximate how wind waves evolve over time. 

This research focuses specifically on spectral wind wave models that are defined through the Wave Action Balance Equation (WAE). These spectral wind wave models work by estimating the transport of wave action distributed among wind waves of various frequencies and directions as defined through the WAE~\cite{komen_cavaleri_donelan_hasselmann_hasselmann_janssen_1994,MONBALIU2003133,Khandekar_1989,Janssen2008,young1999wind}. The WAE presents challenges for numerical discretization techniques because it is a first order transient hyperbolic equation whose domain lies in 4 dimensions (2 in space, frequency, direction) with a non-divergent free velocity field and possibly nonlinear source/sink terms. Numerical spectral wind wave models have been in operation since the 1970s and are used daily as a tool in many fields including engineering, shipping, and meteorology~\cite{TheWAMModel,AThirdGenerationModelforWindWavesonSlowlyVaryingUnsteadyandInhomogeneousDepthsandCurrents,komen_cavaleri_donelan_hasselmann_hasselmann_janssen_1994}. They are also important tools for both improving our understanding of ocean processes and for predicting the impacts of severe weather events. The wind wave models can produce results regarding the state of the sea surface (sea state) such as average wave height, significant wave height, mean weave direction, peak period, and wave radiation stresses.

Currently, there are many well-known spectral wind wave models used in practice such as ECWAM~\cite{TheWAMModel,janssen2004interaction}, which is supported by the European Centre for Medium-Range Weather Forecasts (ECMWF), Simulating WAves Nearshore (SWAN)~\cite{Booij_1999}, which is supported by Delft University of Technology, and WAVEWATCH III~\cite{AThirdGenerationModelforWindWavesonSlowlyVaryingUnsteadyandInhomogeneousDepthsandCurrents}, which is supported by the United States National Oceanic and Atmospheric Administration (NOAA). A brief summary of the methods employed in contemporary numerical spectral follows but we note that there are many great historical reviews on numerical spectral wind wave models see e.g,~\cite{Rolandthesis,Meixnerthesis,komen_cavaleri_donelan_hasselmann_hasselmann_janssen_1994,state_of_art,MONBALIU2003133,Janssen2008,Khandekar_1989,roland2014developments}.

All numerical spectral wind wave models rely on  numerical approximations to the governing partial differential equation (PDE), i.e., the WAE. Historically, spectral wind wave models used finite difference approximations along with splitting schemes to separate the propagation and source terms of the WAE~\cite{state_of_art}. For instance, the ECWAM model contains both a first order explicit upwind scheme and a second order leapfrog method for the propagation terms with a semi-implicit second order centered difference for the source terms~\cite{TheWAMModel}. The original WAVEWATCH III model uses an explicit predictor-corrector method that approximates Crank-Nicholson and an explicit Euler scheme for the source terms~\cite{AThirdGenerationModelforWindWavesonSlowlyVaryingUnsteadyandInhomogeneousDepthsandCurrents}.
The scheme of SWAN was unique at its time of inception in that it used a fully implicit solver without splitting nonlinear and linear terms using the ideas of a nonlinear solver by Patankar~\cite{patankar1980numerical}. In SWAN, a finite difference formulation based on an implicit upwind scheme is used~\cite{Booij_1999}. This idea of using an implicit time step scheme is relatively uncommon in fluid problems such as the WAE but show great promise in applications where the CFL condition is particularly restrictive~\cite{roland2014developments}.

It is also common when splitting the WAE into propagation and source to employ a method of characteristics for the propagation portion~\cite{roland2014developments}. This is an intriguing method because without source terms the WAE is just an advection problem and the characteristics should give the exact solution. However, it can be difficult to use these methods for complex bathymetries and complicated source terms and thus numerical approximations are necessary in practice. Two popular models, STWAVE and TOMAWAC, were developed beginning in the 1990s use this characteristic paradigm and in many cases with great success~\cite{smith2001stwave,benoit1997tomawac}.

As the reliability and usage of wave models increased, there was a push for spectral wind wave models to be used in more shallow waters. Historically, structured grids are common because they are easier to implement and may be computationally less expensive than unstructured grids. However, in order for structured grids to work for coastal applications, very high levels of refinement are needed to capture often highly irregular changes in  bathymetry~\cite{roland2014developments} as well as coastline. Thus, spectral wind wave models capable of unstructured grids became more popular because they are often computationally cheaper than refining a structured grid multiple levels.
 
The first known model capable of computations on unstructured meshes is the aforementioned TOMAWAC~\cite{benoit1997tomawac}, which was released in 1997 and was able to represent the geographic domain with unstructured triangles using what is often known as a characteristic-Galerkin or a Lagrange-Galerkin approach. That is, a method that uses finite element interpolation in combination with a method of characteristics. In~\cite{SWAN_FEM}, Hsu \emph{et al.} developed a finite element method (FEM) version of SWAN that used a Taylor-Galerkin scheme in geographic space while splitting the spectral frequency, direction, and source terms. This model was further developed, focusing on improvements in eliminating oscillations around steep gradients, into wind wave model (WWM) I and WWM II, which added an implicit and explicit fluctuation splitting
schemes (also known as residual distribution schemes) to the geographic space~\cite{roland2006spectral,roland2009development,roland2012application,abgrall2006residual}. This same approach was added into the WAVEWATCH III model, extending WAVEWATCH III to allow for unstructured meshes and implicit time steps~\cite{abdolali2020large}. Around the same time, SWAN became available on unstructured meshes made of triangular elements by extending its finite difference upwinding scheme~\cite{zijlema2010computation}.
 
In addition to the model of Hsu, there have been a couple more finite element methods applied to the WAE. In 2012, Yildirim \emph{et al.} used a hybrid approach with discontinuous Galerkin in geographic space and Fourier collocation in spectral space to create a third generation model~\cite{YILDIRIM20124921}. In 2014, Meixner used a discontinuous Galerkin FEM in spectral and geographic space~\cite{Meixner2014}. Both of the models used Runge-Kutta type explicit time stepping schemes. 

In addition to models that use finite difference methods, and FEMs, there are models that use finite volume methods. FVCOM-SWAVE is an implementation of a finite volume method that allows for unstructured grids~\cite{QI2009153}. A summary of all of the aforementioned schemes are contained in~\textbf{Table~\ref{tab:tab1}}. \markl{As can be seen in~\textbf{Table~\ref{tab:tab1}}, as these operational wind wave models have increased capabilities for capturing the wave spectrum in coastal areas, there are a set of two common features amongst numerical schemes. The first is a capability of fully unstructured triangular grids in geographic space and the second is implicit time stepping. These are important features because unstructured meshes allow for refinement of the irregular bathymetry in the coast and the implicit time step allows for violation of the highly restrictive CFL condition that is inherent in the WAE. This research aims to adhere to these principles but from a finite element perspective.}
\begin{table}[h!]
\begin{tabular}
{|m{2.55cm}|m{1cm}|m{3.25cm}|m{5.0cm}|m{2.5cm}|}
\hline
Model Name & Year & Numerical Method & Method Description & Grid Type \\
\hline

WAM & 1988 & Finite Difference & \textbf{Advection}: Explicit 1st order upwind, leapfrog

\textbf{Source}: Centered difference (semi-implicit)
 & Structured\\
\hline
WAVEWATCH III & 1990 & Finite Difference & \textbf{Advection}: QUICKEST (Explicit 3rd order upwind)

\textbf{Source}: Semi-implicit & Structured\\
\hline
STWAVE & 1990s & Method of Characterstics/ Finite Difference & \textbf{Advection}: Method of characteristics

\textbf{Source}: Finite difference & Structured \\
\hline
TOMAWAC & 1997 & Method of Charactersitcs/FEM & 
\textbf{Advection}: Method of characteristics 

\textbf{Source}: semi-implicit

(linear finite elements for interpolation) & Unstructured\\
\hline

SWAN & 1999,

2008 & Finite Difference & \textbf{Space}: Upwinding

\textbf{Time}: First order implicit & Unstructured\\
\hline
WWMII/WW3 & 2009 & Contour residual scheme & \textbf{Space}: contour residual distribution

\textbf{Time}: Implicit upwind & Unstructured \\
\hline
FVCOM-SWAVE & 2009 & Finite Volume & FVCOM finite volume in space, centered difference in directional, flux corrected transport algorithm for frequency & Unstructured \\
\hline
Yildirim & 2011 & FEM/Spectral & \textbf{Geographic space}: Discontinuous Galerkin

\textbf{Spectral space}: Fourier collocation

\textbf{Time}: TVD Runge-Kutta & Unstructured \\
\hline
Meixner & 2014 & FEM & \textbf{Space}: Discontinuous Galerkin

\textbf{Time}: SSP Runge-Kutta & Unstructured\\
\hline

\end{tabular}
\caption{Summary of numerical schemes in contemporary spectral wind wave models.}\label{tab:tab1}
\end{table}

This work focuses on the development and implementation of a new spectral wind wave model, called WAVEx, that is capable of using unstructured meshes via the FEM along with implicit time stepping using the open source finite element library, FEnicSx~\cite{mcrae2016automated,Habera2020}. FEMs gained popularity starting in the mid twentieth century as a means to numerically approximate differential equations see e.g.,~\cite{zienkiewicz2005finite,becker1981finite,carey1983finite,reddy2019introduction}. In general a FEM consists of recasting a differential equation into an equivalent weak form, partitioning the associated domain into a finite number of subdomains (i.e., elements), approximating the solution on each element with a finite dimensional basis, and subsequently solving the associated linear algebraic system of equations to achieve an approximate solution. 
The FEM has many benefits in that it is backed by rigorous mathematical theory that readily extends to high order approximations, unstructured meshes, and adaptive meshing see e.g.,~\cite{ern2013theory,brenner2008mathematical,zienkiewicz2005finite} for thorough introductions and complete discussions on advantages and weaknesses.  

FEMs were originally developed for elliptic type problems, which often arise in structural mechanics~\cite{brenner2008mathematical,ern2013theory}. However, the standard (Bubnov-Galerkin) FEMs have issues when applied to many common fluid problems, which are often hyperbolic in nature, and has historically led to unstable solutions and solutions that had spurious oscillations~\cite{johnson2012numerical,ern2013theory,reddy2019introduction,oden2017applied}. These problems have been thoroughly analyzed and many stabilized finite element techniques were developed~\cite{johnson1984finite,pironneau1989finite,zienkiewicz2013finite} to address the issues. Some of the methods developed that became popular include artificial viscosity~\cite{ern2013theory}, Taylor-Galerkin~\cite{DONEA1992}, Streamlined Upwind Petrov-Galerkin (SUPG)~\cite{BROOKS1982199}, least squares FEM~\cite{bochev1998finite}, discontinuous Galerkin (DG)~\cite{dawson2001priori,cockburn2000development,brezzi2006stabilization,brezzi2004discontinuous}, and Lagrange-Galerkin or characteristic-Galerkin~\cite{Douglas1982,Bermejo2023,morton1988stability,dawson1994characteristic}. 

In the following study, a set of four finite element formulations of the WAE: SUPG, DG, least squares, and standard Bubnov-Galerkin, will be considered and implemented using the FEniCSx library in a simplified two-dimensional domain (1 in space and 1 in frequency) along with implicit finite difference approximation in time for all FEM schemes. Two test cases will be demonstrated, one of a propagating sinusoidal signal and one with a developing shock. For each test case, the $L^2$ and $l^{\infty}$ errors are calculated and studied with respect to $h$ refinement.   Then, employing the strategies of Loveland \emph{et al.} in~\cite{loveland2022extending}, WAVEx is implemented for the full four dimensional WAE. Thus, WAVEx is a finite element model that employs SUPG stabilization with a second-order implicit time-stepping scheme (Crank-Nicolson) along with a Strang splitting approach to handle source terms in the WAE implemented using FEniCSx. The  open-source framework FEniCSx was chosen, as its powerful open-source framework allows for automation of coding finite element solvers and in hopes that it will allow for users to more easily experiment with different numerical techniques or with different source terms via its Python interface.

Several test cases from the  US Office of Naval Research (ONR) testbed~\cite{ONR_testbed} are reproduced in order to demonstrate the validity and applicability of WAVEx.  The collection of tests include theoretical problems with analytic solutions as well as one lab experiment that has physical measurements. To measure accuracy, the solution generated by WAVEx is compared to the tabulated results at the given points. From these comparison points, both the root mean square error (RMSE), i.e., $l^{2}$ and the $l^{\infty}$ error are computed. For each test case, the parameters include significant wave height and mean wave direction and are computed in WAVEx. \markl{The set of test cases were selected to investigate convergence properties, accuracy, and stability of our numerical schemes as well as demonstrate the scalability of the schemes.} To the knowledge of the authors, no finite element methods have been used for the WAE that use SUPG stabilization, least squares FEM, or discontinuous Galerkin approach along with implicit time stepping. \markl{ Furthermore, to the knowledge of the authors, no full scale wave model has been implemented completely within Python. The use of Python we believe will increase the code's readability as well as ease future collaborations and further development.}

\section{Methods}\label{sec:deriveWAE}
\subsection{Problem Definition}
The Wave Action Balance Equation (WAE) is derived as a conservation law of wave action, $N$, which is a scalar-valued function of horizontal geographical space $(x,y)$ and spectral space $(\sigma,\theta)$, i.e., frequency and direction, and can be found in e.g.,~\cite{holthuijsen_2007,young1999wind,komen_cavaleri_donelan_hasselmann_hasselmann_janssen_1994,leblond1981waves}. Assuming source terms are linear with respect to $N$,  the WAE is a linear, scalar-valued, hyperbolic equation in 4 dimensions with a varying (both in 4-D space and time), non-divergence free velocity field,  ($\mathbf{c}(x,y,\sigma,\theta)$), which can be determined independently of the unknown $N(x,y,\sigma,\theta)$. This is different than most conservation laws such as Navier-Stokes and related transport equations because typically the propagation velocity is just a function of the unknown i.e. $\mathbf{c}(N)$. 

The proper boundary and initial conditions are as follows, assuming we have a bounded domain in four-dimensional space $\Omega \subset \mathbb{R}^4$ that is sufficiently regular. Similar to other advection problems, the boundary is split up into 2 segments, inflow and outflow:
\begin{equation}
    \begin{split}
        \Gamma_-={x\in \partial \Omega : \mathbf{c} \cdot \mathbf{n} < 0} = \textrm{inflow} \\
        \Gamma_+={x\in \partial \Omega : \mathbf{c} \cdot \mathbf{n} \geq 0} = \textrm{outflow} ,
    \end{split}
\end{equation}
where $\mathbf{n}$ is the outwards unit normal vector to the boundary. Then it has been shown that the following problem possesses a unique solution~\cite{leveque1992numerical,DONEA1992,pironneau1989finite}:
\begin{equation}\label{eqn:WAE}
    \begin{split}
        N_t + \nabla \cdot (\mathbf{c}N) = \frac{S(N,x,y,\sigma,\theta,t)}{\sigma} \quad \textrm{on} \quad \Omega \times (0,T),\\
        N = N_- \quad \mathrm{on} \quad \Gamma_-,\\
        N = N_0 \quad \textrm{on} \quad \Omega \quad \textrm{at} \quad t =0,
    \end{split}
\end{equation}
where $N_t$ denotes the (partial) time derivative of $N$, $S$ the source/sink terms, $\Omega$ the computational domain, and $N_-$ the specified essential boundary condition on the outflow boundary.
In the case when $S$ is either $0$ or is independent of $N$, a fairly straightforward analytic solution to the above problem can be obtained via the method of characteristics~\cite{DONEA1992}. However, in practice $S$ is quite complex, often nonlinear, and realistic problems become too complex to analyze with pencil and paper and thus the need for numerical methods to approximately solve~\eqref{eqn:WAE} arises. 

The propagation velocities, $\mathbf{c}$, are all determined using the constitutive relation called the dispersion relation, which is derived under assumptions of Airy Wave Theory. The dispersion relation relates relative radial frequency, $\sigma$, to wavenumber magnitude, $k$: 
\begin{equation}\label{eqn:dispersion}
    \sigma^2=gk \tanh(kd).
\end{equation}
Here $g$ is the constant of gravitational acceleration and $d$ is total water depth. The velocity $\mathbf{c}$ is a non-constant, non-divergence free vector quantity defined as:
\begin{equation}
    \mathbf{c} = (c_g cos \theta + u, c_g sin \theta + v, c_{\sigma}, c_{\theta} ),
\end{equation}
where $u,v$ are the mean water velocities in $x, y$ directions respectively,  $c_g$ the relative group velocity  (defined as $\frac{\partial \sigma}{\partial k}$, whereas the absolute group velocity is $\frac{\partial \omega}{\partial k} = c_g + \|(u,v)\|$), and $c_{\sigma}$, $c_{\theta}$ define the advection of the spectra with respect to direction and relative frequency, respectively. The group velocity, $c_g$, can be directly obtained by differentiating the dispersion relation in~\eqref{eqn:dispersion} for relative frequency, $\sigma$, with respect to wavenumber magnitude $k$. The explicit expression is:
\begin{equation}\label{eqn:groupveloc}
    c_g = \frac{1}{2}\left( 1 + \frac{2kd}{\sinh{2kd}}\right ) \sqrt{\frac{g}{k} \tanh{(kd)}}.
\end{equation}
The propagation velocity $c_{\sigma}$, represents frequency shifting due to changes in depths and currents. By applying the chain rule, we can find an expression of $c_{\sigma}$: 
\begin{equation}
\begin{split}
        c_{\sigma} = \frac{d \sigma}{d t} =& \frac{k \sigma }{\sinh{(2kd)}}\left( \frac{\partial d}{\partial t} + u \frac{\partial d}{\partial x} + v\frac{\partial d}{\partial y}\right) - \\
        &c_g k\left( \frac{\partial u}{\partial x}\text{cos}^2(\theta) + \frac{\partial u}{\partial y}\text{cos}(\theta)\text{sin}(\theta) + \frac{\partial v}{\partial x}\text{sin}(\theta)\text{cos}(\theta) + \frac{\partial v}{\partial y}\text{cos}^2(\theta)\right).
\end{split}
\end{equation}
The full derivation is quite lengthy but can be found in Appendix D in~\cite{holthuijsen_2007}.  Now the propagation velocity $c_{\theta}$, with respect to $\theta$,  represents the shift in the spectrum due to refraction and diffraction. Similarly to $c_{\sigma}$, the derivation is quite lengthy but can be obtained by applying the chain rule:
\begin{equation}
\begin{split}
       c_{\theta}=\frac{d \theta}{d t} =& \frac{\sigma}{\sinh{(2kd)}} \left(  \frac{\partial d}{\partial x}\text{sin}(\theta) - \frac{\partial d}{\partial y}\text{cos}(\theta)\right ) + \\
       &\frac{\partial u}{\partial x}\text{cos}(\theta)\text{sin}(\theta) - \frac{\partial u}{\partial y}\text{cos}^2(\theta) + \frac{\partial v}{\partial x}\text{sin}^2(\theta) - \frac{\partial v}{\partial y}\text{cos}(\theta)\text{sin}(\theta).
\end{split}
\end{equation}

\subsection{Source Terms}\label{chap:SourceTerms}
In the WAE~\eqref{eqn:WAE}, the source term $S(N,x,y,\sigma,\theta,t)$ was left arbitrary. However, in practice,   it can take many forms but in general it can be thought of as a sum of three key sources/sinks:
\begin{equation}\label{eqn:Hass}
    S = S_{in} + S_{diss} + S_{nl}.
\end{equation}
where $S_{in}$ represents any contribution to the spectrum due to wind input, $S_{diss}$ represents any change in the energy spectrum due to dissipation that can include things like whitecapping and surf zone breaking, and $S_{nl}$ represents changes in the spectrum due to nonlinear wave interactions. When source terms are active, the WAVEx model in this study will use the default 3rd generation source term package used in SWAN cycle III Version 41.41, see details in~\cite{swantech}. The default wind input term is that from WAM Cycle III and first outlined in~\cite{Komen1984}, the dissipation from whitecapping, bottom friction, and depth-induced breaking are included and the DIA approximation is used to account for nonlinear interactions. More details on these source terms can be found in  Chapter 3 of~\cite{markthesis}.

\subsection{Weak Formulations For the Wave Action Balance Equation}\label{chap:stabilized_methods}

In this section, several weak forms of the WAE will be presented. These weak forms will eventually lead to to fully discrete systems that are then solved with FEniCSx. First, for comparison purposes, two forms of the standard Bubnov-Galerkin approach will be derived for the WAE and the problems with the standard Bubnov-Galerkin discretizations will be discussed. Three alternatives to the standard Bubnov-Galerkin approach: the least squares FEM, the SUPG method, and the DG method will be derived and discussed. All of these weak forms will then be run on two simplified test cases using FEniCSx in Section~\ref{sec:convergence} in order to investigate convergence properties of all the stabilized finite element methods with respect to $h$ refinement. The full discretization of an SUPG method will be described for the full four-dimensional setting in the subsequent section. Note that we utilize inner product notation for integrals, i.e., $\left(u ,v\right)_{\Omega} = \int_\Omega u v \, \rm{d}\Omega$.

\subsubsection{Bubnov-Galerkin}

Taking the strong form of the WAE as in~\eqref{eqn:WAE}, we obtain an equivalent problem but in the weak form by multiplying both sides by a test function $w \in L^2(\Omega)$ and integrating over a bounded and Lipschitz domain $\Omega \subset \mathbb{R}^4$. Note that we do not integrate over time because we will eventually employ finite difference approximation in this dimension. Thus we have the following weak form that we will call CG Strong because there are more regularity requirements on the trial space. Find $N \in U$ such that:
\begin{equation}\label{eqn:Galerkin1}
        \left(\frac{\partial N}{\partial t}  ,w\right)_{\Omega} + \left(\nabla \cdot (\textbf{c} N),w\right)_{\Omega} = \left(\frac{S}{\sigma},w\right)_{\Omega} \quad \forall w \in L^2(\Omega).
\end{equation}
Thus, it is required that $U = \{N \in H^1({\Omega},t); N(\Gamma_-,t) = N_-; \frac{\partial N}{\partial t} \in L^2(\Omega,t) \}$, recalling that $\Gamma_-$ is the inflow boundary and $N_-$ is the Dirichlet condition on that part of the boundary. We will also assume that velocity vector $\mathbf{c}$ and source term $S$ is sufficiently well behaved for the weak form to have meaning. It can be shown that~\eqref{eqn:Galerkin1} is well-posed (in the sense that a solution exists, it is unique, and continuously depends on the data) see e.g., the book of Ern and Guermond~\cite{ern2013theory}. However, it is also known that if~\eqref{eqn:Galerkin1} is discretized using the standard Bubnov-Galerkin approach then the discrete inf-sup condition depends on mesh size, $h$, and therefore doesn't provide optimal error estimates~\cite{ern2013theory}. By optimal, we mean for polynomial approximations of order $k$ the convergence of the error in the $L^2$ sense to of the same order of  convergence as the underlying FE interpolants, usually $h^{k+1}$ (this classical result can be found in e.g.,~\cite{babuska2010finite,brenner2008mathematical,reddy2019introduction,oden2017applied}). In practice this leads to oscillatory solutions because only $L^2$ stability can be shown~\cite{johnson1984finite,zienkiewicz2005finite}. Also, the standard Bubnov-Galerkin approximation in this setting often leads to ill-conditioned systems of equations and can even lead to divergence in the finite element solution especially if the exact solution is non-smooth~\cite{zienkiewicz2005finite,johnson2012numerical}. 

We can also derive a second weak form if we integrate by parts, which we will call the CG form, find $N \in U$:
\begin{equation}\label{eqn:Galerkin}
            \left(\frac{\partial N}{\partial t}  ,w\right)_{\Omega} - \left( \textbf{c} N,\nabla w\right)_{\Omega} + \left( \mathbf{c} N \cdot \mathbf{n},w\right)_{\Gamma_+}= \left(\frac{S}{\sigma},w\right)_{\Omega} \quad \forall w \in H^1(\Omega). 
\end{equation}
In order for the term on the boundary to have meaning we must require additional regularity than just $N \in L^2(\Omega)$ and so the analysis is not so straightforward but has been shown to be well posed for cases of homogeneous boundary conditions and in cases with lower dimensionality~\cite{ern2021guaranteed}. When discretized, similar problems can be expected to that of the strong CG from in~\eqref{eqn:Galerkin1}. In order to mitigate the problems inherent in these standard Bubnov-Galerkin discretizations, we will investigate a least squares method, an SUPG method, and a DG method.

\subsubsection{Least Squares}
The least squares approach is fundamentally different to the Galerkin approach in that the least squares method is posed as a minimization problem of a linear differential operator. A good reference on the properties of least squares can be found in the text of Bochev and Gunzburger~\cite{FEMLSQ_book}. To derive the least squares form for the WAE, we start with the strong form as in~\eqref{eqn:WAE},
we define the differential operator $\mathcal{L}$ and right hand side as in~\cite{FEMLSQ_book} so that the above equation $\mathcal{L}$ would be defined:
\begin{equation}
    \mathcal{L}\bullet =  \frac{\partial \bullet}{\partial t} + \nabla \cdot (\textbf{c} \bullet).
\end{equation}
Then the least squares variational form then is defined as follows. Find $N \in U$:
\begin{equation}
    (\mathcal{L}N, \mathcal{L} w)_{\Omega} = (\frac{S}{\sigma},\mathcal{L} w)_{\Omega} \quad \forall w \in U.
\end{equation}
This problem is equivalent to minimizing the functional $J(V) = \frac{1}{2}(\mathcal{L}(V),\mathcal{L}(V))_{\Omega} - (\mathcal{L}V,S/\sigma)_{\Omega}$. Explicitly writing out the definition of $\mathcal{L}$ gives the formulation for the WAE will be, find $N \in U$:
\begin{equation}
    (\frac{\partial N}{\partial t} + \nabla \cdot (\textbf{c} N), \nabla \cdot (\textbf{c} w))_{\Omega} = (\frac{S}{\sigma},\nabla \cdot (\textbf{c} w))_{\Omega} \quad \forall w \in U,
\end{equation}
where $U = \{N \in H^1(\Omega,t); N(\Gamma_-,t) = N_-; \frac{\partial N}{\partial t} \in L^2(\Omega) \} $, and again $\mathbf{c}, S$ are assumed sufficiently regular to give the integrals meaning. We note that the time derivative on the test space will be omitted because the time derivative will be discretized via finite difference and not finite elements.

The least squares problem generates a symmetric problem (trial and test spaces are the same) and thus when discretized exhibits different properties than the Bubnov-Galerkin approximations. It has been shown that for similar problems, such as the advection-reaction equation, the least squares method can reduce oscillations and provide stability~\cite{ern2013theory}. It can be shown that least squares produces optimal error convergence in the streamline directions however it has yet to be shown that the least squares method produces optimal errors with respect to $h$ refinement in the $L^2$ norm~\cite{ern2013theory}.

\subsubsection{SUPG}
The next stabilized form that will be investigated is the SUPG method. Because the WAE is an advection dominated equation, the method is a natural choice to employ. The method originally from~\cite{BROOKS1982199} was designed for fluid problems and has had much success in stabilizing advection dominated problems for fluids by adding diffusion in the upwind direction in the weak formulation. The SUPG scheme is formulated by adding a term with a vanishing residual to the standard Bubnov-Galerkin form in the discrete section. This in turn modifies the discrete test space, which results in the change from a standard Bubnov-Galerkin form to a so called Petrov-Galerkin form. We can add this residual term to either CG strong as in~\eqref{eqn:Galerkin1} or the CG form as in~\eqref{eqn:Galerkin}. We will demonstrate both possible SUPG formulations now. We start with CG strong form and taking it into the discrete setting where the test and trial functions are piecewise smooth and globally continuous. We will call the resulting form SUPG strong, which is find $N_h \in U_h(\Omega)$:
\begin{equation}
            \left(\frac{\partial N_h}{\partial t} + \nabla \cdot (\textbf{c} N_h),w_h\right)_{\Omega} + \left(\frac{\partial N_h}{\partial t} +  \nabla \cdot (\textbf{c} N_h )- \frac{S}{\sigma}
            ,  \tau \mathbf{c} \cdot  \nabla (w_h)\right)_{\Omega_e} = \left(\frac{S}{\sigma},w_h\right)_{\Omega} \quad \forall w_h \in V_h(\Omega).
\end{equation}
Here $\Omega_e$ is the interior over each finite element. Because in the analytic setting the residual term is identically 0, then this problem inherits the well-posedness from~\eqref{eqn:Galerkin1}. The weak form can can be simplified by adding the integrals together:
\begin{equation}\label{eqn:SUPG_I}
             \left(\frac{\partial N_h}{\partial t} + \nabla \cdot (\textbf{c} N_h )
            , w_h + \tau  \mathbf{c} \cdot \nabla  (w_h)\right)_{\Omega_e} = \left(\frac{S}{\sigma},w_h + \tau \mathbf{c} \cdot \nabla  (w_h)\right)_{\Omega_e} \quad \forall w_h \in V_h(\Omega).
\end{equation}
This will be called the SUPG strong form that will be discretized with continuous finite elements in FEniCSx. It has been shown that adding this diffusion in the streamline direction improves the stability of the discrete problem by improving the discrete inf-sup condition (see~\cite{johnson1984finite,BROOKS1982199}) and allowing for discrete stability in the $H^1$ sense. Furthermore, error estimates for very similar problems to the WAE, such as the advection-reaction equation, show nearly optimal rates at $O(h^{k+0.5})$~\cite{johnson1984finite,burman2010consistent,burman2011analysis}. 

We can also arrive at an SUPG form by taking~\eqref{eqn:Galerkin} into the discrete setting and adding the residual to get, find $N_h \in U_h(\Omega)$:
\begin{equation}\label{eqn:SUPGII}
    \begin{split}
                \left(\frac{\partial N_h}{\partial t}  ,w_h\right)_{\Omega} - \left( \textbf{c} N_h,\nabla w_h\right)_{\Omega} + \left( \mathbf{c} N_h \cdot \mathbf{n},w_h\right)_{\Gamma_+} + \\ \left(\frac{\partial N_h}{\partial t} + \nabla \cdot (\textbf{c} N_h )- \frac{S}{\sigma}
            , \tau\mathbf{c} \cdot \nabla (w_h)\right)_{\Omega_e}= \left(\frac{S}{\sigma},w_h\right)_{\Omega}  \quad \forall w_h\in V_h(\Omega).
    \end{split}
\end{equation}
After descritization of both the SUPG strong and SUPG forms, $\tau$ is taken from \cite{CODINA200061,burman2010consistent} and is chosen to be:
\begin{equation}
    \tau = \frac{h}{\|\mathbf{c}\|},
\end{equation}
where $h$ is the characteristic length of an element.

\subsubsection{Discontinuous Galerkin}
The next stabilized form that will be examined is the DG method as in the work of Meixner~\cite{Meixner2014}. The DG method is different to any of the other methods discussed up to this point in that it allows for discontinuities between element interfaces.  The DG formulation is obtained by taking the strong form of the WAE, multiplying by a test function, and integrating by parts over each element. Find $N_h \in U_h$:
\begin{equation}
            \left(\frac{\partial N_h}{\partial t}  ,w_h\right)_{\Omega_e} - \left( \textbf{c} N_h,\nabla w_h\right)_{\Omega_e} + \left( \mathbf{c} \hat{N}_h \cdot \mathbf{n},w_h\right)_{\partial\Omega_e}= \left(\frac{S}{\sigma},w_h\right)_{\Omega_e}   \quad \forall w_h \in W_h.
\end{equation}
The test and trial spaces are virtually the same with $W_h = \{v\in L^2(\Omega) : v|_{\Omega_e} \in P^k(\Omega_e) \, \forall \Omega_e \in T_h\}$ and $U_h$ being $W_h$ but with a finite energy lift on the inflow boundary. $P^k$ are polynomials of degree $k$, and $\Omega_e$ corresponds to the interior of an element while $T_h$ is s finite element partition of the domain $\Omega$. $\hat{N}_h$ is a uniquely defined flux on the boundary, which is required to form a closed system of equations (see~\cite{Meixner2014,cockburn2000development,dawson2001priori}).  In this implementation, the fluxes on the element boundaries are uniquely defined in this case using the Lax-Friedrichs flux, which is described in detail in the work of Meixner \emph{et al.} 2014~\cite{Meixner2014}. The DG method has well documented analysis on stability and error estimation for hyperbolic problems such as the advection reaction equation, which is essentially what the Wave Action Balance Equation is~\cite{cockburn2000development,brezzi2004discontinuous,brezzi2006stabilization}. It has been shown that for a DG scheme like the one above, stability in the $H^1$ sense exists and that almost optimal convergence, $O(h^{k+\frac{1}{2}})$ can be shown~\cite{brezzi2004discontinuous}.

\subsection{WAVEx: Implementation of Wave Action Balance Equation Solver with FEniCSx}\label{chap:WAVexAcBal}

We now introduce WAVEx, which is capable of solving the WAE in its full 4-dimensional spatial domains using the techniques described in~\cite{loveland2022extending}, and is implemented with the finite element library FEniCSx. WAVEx in its current state allows for SUPG stabilization along with Strang splitting to deal with the nonlinear source terms. The codebase for WAVEx is publically available and on Github at~\href{https://github.com/Markloveland/WAVEx}{https://github.com/Markloveland/WAVEx}. This section will explain in detail the numerics currently available in WAVEx including the finite element schemes as well as the time stepping schemes.

\subsubsection{Semi-Discrete Form}
 This particular implementation in WAVEx will use a continuous FEM approach along with SUPG stabilization. We first limit ourselves to cases where the global domain is a Cartesian product between two subdomains (geographic space, and spectral space). This means we can construct the product basis:
\begin{equation}
    N(x,y,\sigma,\theta) \approx \sum_{i=1}^N \sum_{j=1}^M N_{i,j} \phi_i(x,y) \psi_j(\sigma,\theta),
\end{equation}
where $N_{i,j}$ are the scalar weights of the interpolating functions.

As we know, the standard Bubnov-Galerkin approximation of advective equations leads to ill conditioned discrete problems and as in Section~\ref{chap:stabilized_methods}, we will use SUPG stabilization. SUPG was chosen because it performed well in our preliminary 2D studies (see Section~\ref{sec:convergence}), and the fact that it is also computationally cheapest. We begin by first recalling the two possible weak forms as described in Section \ref{chap:stabilized_methods}, SUPG strong and SUPG. The latter being when integration by parts is conducted and the former is when integrating by parts is omitted. Both options are included in the  WAVEx code repository. For the form without integration by parts, SUPG strong, we begin with~\eqref{eqn:SUPG_I} and substitute in the product basis to get the semi-discrete (i.e., discrete in space, infinite dimensional in time) system:
\begin{equation}\label{eqn:semi-SUPGI}
    (\frac{\partial (\phi_i \psi_j N_{i,j})}{\partial t} + \nabla \cdot (\mathbf{c} \phi_i\psi_j N_{i,j}) - \phi_i \psi_j S_{i,j}, \gamma_k \beta_l + \tau \mathbf{c}  \cdot \nabla (\gamma_k\beta_l) )_{\Omega}= 0.
\end{equation}
The source terms, $S$ in general are quite complex, so we will only seek the $L^2$ projections of their exact form as shown in~\eqref{eqn:semi-SUPGI}.
Analogously, the SUPG weak form of the WAE~\eqref{eqn:SUPGII} in the semidiscrete setting  is defined by substituting in our product basis into~\eqref{eqn:SUPGII} to get:
\begin{equation}\label{eqn:semi-SUPGII}
\begin{split}
    (\frac{\partial (\phi_i \psi_j N_{i,j})}{\partial t},\gamma_k \beta_l)_{\Omega} - (\mathbf{c} \phi_i \psi_j N_{i,j}, \nabla (\gamma_k\beta_l))_{\Omega} + (\mathbf{c}\phi_i\psi_j \cdot \mathbf{n} N_{i,j},\gamma_k\beta_l)_{\partial \Omega} + \\
    (\frac{\partial (\phi_i \psi_j N_{i,j})}{\partial t} + \nabla \cdot (\mathbf{c} \phi_i\psi_j N_{i,j}) - \phi_i \psi_j S_{i,j}, \tau \mathbf{c} \cdot \nabla ( \gamma_k\beta_l) )_{\Omega}= (\phi_i\psi_j S_{i,j},\gamma_k\beta_l)_{\Omega}.
\end{split}
\end{equation}
This takes care of the terms not related to the time step, the choice of finite difference approximation in time will dictate the full discretization.

\subsubsection{Time Step Choice and Fully Discrete Form}\label{sec:timestep}
WAVEx uses a generalized one step finite difference approximation for the time derivative in~\eqref{eqn:semi-SUPGI} and~\eqref{eqn:semi-SUPGII}:
\begin{equation}\label{eqn:fulldiscrete}
\begin{split}
    (\frac{\partial(\phi_i\psi_jN_{i,j})}{\partial t},\gamma_k\beta_l + \tau \mathbf{c} \cdot \nabla  (\gamma_k\beta_l ))_{\Omega} = F(N,t) \rightarrow \\
     (\frac{\phi_i^{n+1}\psi_j^{n+1}N_{i,j}^{n+1} - \phi_i^n\psi_j^n N_{i,j}^n}{\Delta t}, \gamma_k\beta_l + \tau \mathbf{c} \cdot \nabla (\gamma_k\beta_l ))_{\Omega} = \alpha_t F(N,t^{n+1}) + (1-\alpha_t) F(N,t^{n}).
\end{split}
\end{equation}
where term $F(N,t)$ is shorthand from~\eqref{eqn:semi-SUPGI} and equals:
\begin{equation}\label{eqn:flux1}
    F(N,t)=  -(\nabla \cdot (\mathbf{c} \phi_i\psi_j N_{i,j}) - \phi_i \psi_j S_{i,j}, \gamma_k \beta_l + \tau \mathbf{c} \cdot \nabla( \gamma_k\beta_l) )_{\Omega},
\end{equation}
and for~\eqref{eqn:semi-SUPGII}: 
\begin{equation}\label{eqn:flux2}
\begin{split}
        F(N,t) =  (\mathbf{c} \phi_i \psi_j N_{i,j}, \nabla (\gamma_k\beta_l))_{\Omega} - (\mathbf{c}\phi_i\psi_j \cdot \mathbf{n} N_{i,j},\gamma_k\beta_l)_{\partial \Omega} + (\phi_i\psi_j S_{i,j},\gamma_k\beta_l)_{\Omega} \\  - (\nabla \cdot (\mathbf{c} \phi_i\psi_j N_{i,j}) - \phi_i \psi_j S_{i,j}, \tau \mathbf{c} \cdot \nabla( \gamma_k\beta_l) )_{\Omega},
\end{split}
\end{equation}
the superscripts represent the discrete level in time, $n$ will be the previous time step and $n+1$ is the point in time to be solved for, and $\alpha_t$ is a time step parameter between $0$ to $1$. The scheme is implicit unless $\alpha_t =0$, as $\alpha_t =0$ would be the first order explicit Euler approximation, $\alpha_t=1$ would be the first order implicit Euler approximation, and $\theta=0.5$ would be the second order Crank-Nicolson scheme. The terms with superscript $n$ are known and move to the RHS while the terms $n+1$ will be part of the global stiffness matrix at each time step.

The definition given for the finite difference approximation to the time derivative leads to a fully discrete system described by~\eqref{eqn:fulldiscrete} depending on the choice of $F(N,t)$. Choosing~\eqref{eqn:flux1} will yield the fully discrete system for SUPG strong (no integration by parts) and choosing~\eqref{eqn:flux2} will give the fully discrete system for SUPG weak. Once a choice of finite basis is made for trial and test spaces, the forms above will yield a system of equations that in turn can be solved for the action balance at the next time step, $N^{n+1}_{i,j}$. However, there are two difficulties that need to be addressed in order to reach a final implementation: $i)$, FEniCS(x) can not evaluate these integrals on the entire 4-dimensional domain $\Omega$. In the next section, the discrete equations in~\eqref{eqn:fulldiscrete} will be broken down into products of the geographic and spectral subdomains using the product rule as in~\cite{loveland2022extending} into a discrete form that is implementable in FEniCSx.  $ii)$, in general the source term $S$ is nonlinear with respect to $N$. This means that the discrete form of~\eqref{eqn:fulldiscrete} will not yield a system of linear equations as long as we are using an implicit time step, $\alpha_t > 0$. To avoid this, an operator splitting method will described in Section~\ref{sec:strang_split} in order to avoid having to solve a nonlinear system. Although in the future a fully implicit solver for the nonlinear problem is of great interest.

\subsubsection{FEniCSx Compatible Format}\label{sec:product_rule}
Now because we want to only evaluate functions in the 2-dimensional subdomains with FEniCSx we need to take advantage of the product rule. As an example, starting with the second term from~\eqref{eqn:semi-SUPGII} we get:
\begin{equation}
 - (\mathbf{c} \phi_i \psi_j, \nabla (\gamma_k\beta_l))_{\Omega} =  - (\mathbf{c} \phi_i \psi_j, \gamma_k \nabla (\beta_l) + \beta_l \nabla (\gamma_k)  )_{\Omega}. 
\end{equation}
Similar to the previous examples, using the property that the basis functions are only non-constant in their respective subdomains we get the following simplifications ($\nabla_1$ denotes gradient operator in the geographic subdomain, and $\nabla_2$ denotes the gradient operator in the spectral subdomain, and same for velocity):
\begin{equation}
    - (\mathbf{c} \phi_i \psi_j, \gamma_k \nabla (\beta_l) + \beta_l \nabla (\gamma_k)  )_{\Omega} =  - (\mathbf{c}_1 \phi_i \psi_j,\beta_l \nabla_1 (\gamma_k))_{\Omega} - (\mathbf{c}_2 \phi_i \psi_j, \gamma_k \nabla_2 (\beta_l)  )_{\Omega}.
\end{equation}
This importantly allows us to pull integrands apart in the following fashion so that we can use FEniCSx:
\begin{equation}\label{eq:SUPGII_term1}
\begin{split}
         - (\mathbf{c}_1 \phi_i \psi_j,\beta_l \nabla_1 (\gamma_k))_{\Omega} - (\mathbf{c}_2 \phi_i \psi_j, \gamma_k \nabla_2 (\beta_l)  )_{\Omega} = -\int_{\Omega_2} \psi_j \beta_l \int_{\Omega_1} \phi_i \mathbf{c}_1 \cdot \nabla_1 \gamma_k dxdy \\
         - \int_{\Omega_2} \psi_j\nabla_2\beta_l \cdot \int_{\Omega_1} \mathbf{c}_2 \phi_i \gamma_k dxdy.
\end{split}
\end{equation}
Where~\eqref{eq:SUPGII_term1} represents a term that can be evaluated using FEniCSx. Integrands inside $\Omega_1$ each create a $N_{dof1} \times N_{dof1}$ sparse matrix and to form the full discrete system we will need a matrix at each quadrature point in the second subdomain. Each matrix will be different because of the changes in $\mathbf{c}$ in both subdomains. The remainder of the terms in ~\eqref{eqn:fulldiscrete} can be rewritten in this way via the product rule. The full process is the same can be seen in detail in Appendix~\ref{ap:B}.

\subsubsection{Operator Splitting}\label{sec:strang_split}
The source terms are in general non-linear, in the interest of saving computational complexity we will employ a splitting scheme to handle the source terms of the WAE. More extensive notes on operator splitting can be found in ~\cite{Glowinski2016}. We will be employing the second order Strang splitting~\cite{StrangSplitting1968} scheme, which can be defined on a system of ODEs as follows:

%
%
\begin{equation}
    \begin{split}
        \frac{dX}{dt} + A(X,t) = 0, \in (0,T),\\
        X(0) = X_0.\\
    \end{split}
\end{equation}
In our case $A$ will be the discrete finite element operator in 4D space and $X$ will be the action balance at all of the degrees of freedom at a particular point in time. The Strang splitting scheme splits up $A$ as a sum of two operators $A_1$ and $A_2$ for $n \geq 0,X^n \rightarrow X^{n+\frac{1}{2}} \rightarrow \hat{X}^{n+\frac{1}{2}} \rightarrow X^{n+1} $. Solve:
%
%
%
%
%
%
\begin{equation}
\begin{split}
    \frac{dX_1}{dt} + A_1(X_1,t) = 0 , t\in (t^n,t^n+\frac{1}{2}),\\
    X_1(t^n) = X^n,\\
\end{split}
\end{equation}
then set $X^{n+\frac{1}{2}} = X_1(t^{n+\frac{1}{2}})$ and solve: 
\begin{equation}
\begin{split}
        \frac{dX_2}{dt} + A_2(X_2,t) = 0, t\in(t^n,t^n+\Delta t),\\
    X_2(t^n) = X^{n+\frac{1}{2}},\\
\end{split}
\end{equation}
then set $\hat{X}^{n+\frac{1}{2}} = X_2(\Delta t)$ and solve:
\begin{equation}
    \begin{split}
        \frac{dX_1}{dt} + A_1(X_1,t) = 0, t \in (t^{n+\frac{1}{2}},t^{n+1}), \\
        X_1(t^{n+\frac{1}{2}}) = \hat{X}^{n+\frac{1}{2}}.\\
    \end{split}
\end{equation}
This gives $X^{n+1} = X_1(t^{n+1})$. We will apply this second order Strang splitting to the WAE by setting $A_1$ to be the source terms and $A_2$ as all remaining terms. In particular, for the case of the semi-discretization in~\eqref{eqn:semi-SUPGI}, the operators will be:
%
%
%
%
\begin{equation}
    \begin{split}
        A_1 = -(\phi_i \psi_j S_{i,j}, \gamma_k \beta_l + \tau \mathbf{c}  \cdot \nabla (\gamma_k\beta_l) )_{\Omega}, \\
        A_2 = ( \nabla \cdot (\mathbf{c} \phi_i\psi_j N_{i,j}), \gamma_k \beta_l + \tau \mathbf{c}  \cdot \nabla (\gamma_k\beta_l) )_{\Omega}.
    \end{split}
\end{equation}
We will implement the Strang splitting with an implicit-explicit scheme. The time step associated with operator $A_2$ will be the scheme from Section~\ref{sec:timestep}, whereas the two time steps associated with operator $A_1$ will be an explicit second order Runge-Kutta (RK2) scheme~\cite{GottliebSSPRK2018}. The RK2 used in WAVEx is defined as:
\begin{equation}
    \begin{split}
        N^{(1)} = N^{n} + \Delta t A_1(N^n,t^n),\\
        N^{n+1} = \frac{1}{2}N^{n} +  \frac{1}{2}(N^{(1)} + \Delta t A_1(N^{(1)}, t^{n+1})).
    \end{split}
\end{equation}
%
%

To summarize, WAVEx utilizes SUPG stabilization in 4 dimensions and employs Strang splitting to separate advection from the source terms. The source terms are integrated in time with an explicit RK2 scheme while the advection part is integrated in time with an implicit generalized one step finite difference. Because WAVEx uses the FEniCSx framework, it is automatically parallelized via MPI and the linear algebra backend that is used is the Portable, Extensible Toolkit for Scientific Computation (PETSc)~\cite{balay2019petsc}. The linear solver that is used in all the following cases is the default GMRES solver with block Jacobi preconditioning. Also, WAVEx is designed to allow for the addition of different discretizations in both time and space besides just the specific choices made in this case. 

\section{Convergence Test Case Results}\label{sec:convergence}
To investigate the asymptotic convergence properties of the stabilized methods introduced in Section~\ref{chap:stabilized_methods}, two test cases will be considered. These test cases will be run with reduced dimension (2 instead of 4 in space) and the velocity fields will be simplified. For each case the error in the finite element solution, $N_h$, relative to the exact solution, $N_{exact}$, will be computed both in the $L^2(\Omega)$ and the $l^{\infty}(\Omega)$ norm that are defined as:
\begin{equation}\label{eqn:L2Linf}
    \begin{split}
        \| N_h-N_{exact} \|_{L^2} = \sqrt{\int_{\Omega} (N_h - N_{exact} )^2 dx},  \\
        \| N_h - N_{exact} \|_{l^{\infty}} = \max_{i \in n_{dof}}  ( | N_h(x_i) -N_{exact}(x_i) | ) .
    \end{split}
\end{equation}

\subsection{Propagation of a Sinusoidal Wave}
The first test case is a harmonic wave propagating through a square domain over time. The goal of this test case is to evaluate the capabilities of the different FEM schemes in a "pure advection" scenario by setting source, $S$, to 0. Harmonic waves are propagating across 2D domain in the direction of the velocity vector, $\mathbf{c} = (1,1)$, which is constant in space and time. The domain is a square, $\Omega = (0,10) \times (0,10)$.  The exact solution is:
\begin{equation}
    N_{analytic} = sin(x - c_xt) + cos(y - c_yt).
\end{equation}
This dictates the initial and boundary conditions:
\begin{equation}
    \begin{split}
        N_0(x,y) = sin(x) + cos (y), \\
        N(x,y,t) = N_{analytic} \quad \textrm{for } (x,y) \in \Gamma_-,
    \end{split}
\end{equation}
and as a consequence of the velocity field the Dirichlet boundary, $\partial\Omega_d$, is the bottom and left sides of the square domain. The test case has a total simulation time of 5 units, which is split up into 1000 time steps. The time step is set small to minimize error from the approximation in time and focus on FEM error convergence with respect to $h$ refinement. 

\begin{figure}[h!]
    \centering
    \includegraphics[width=\textwidth]{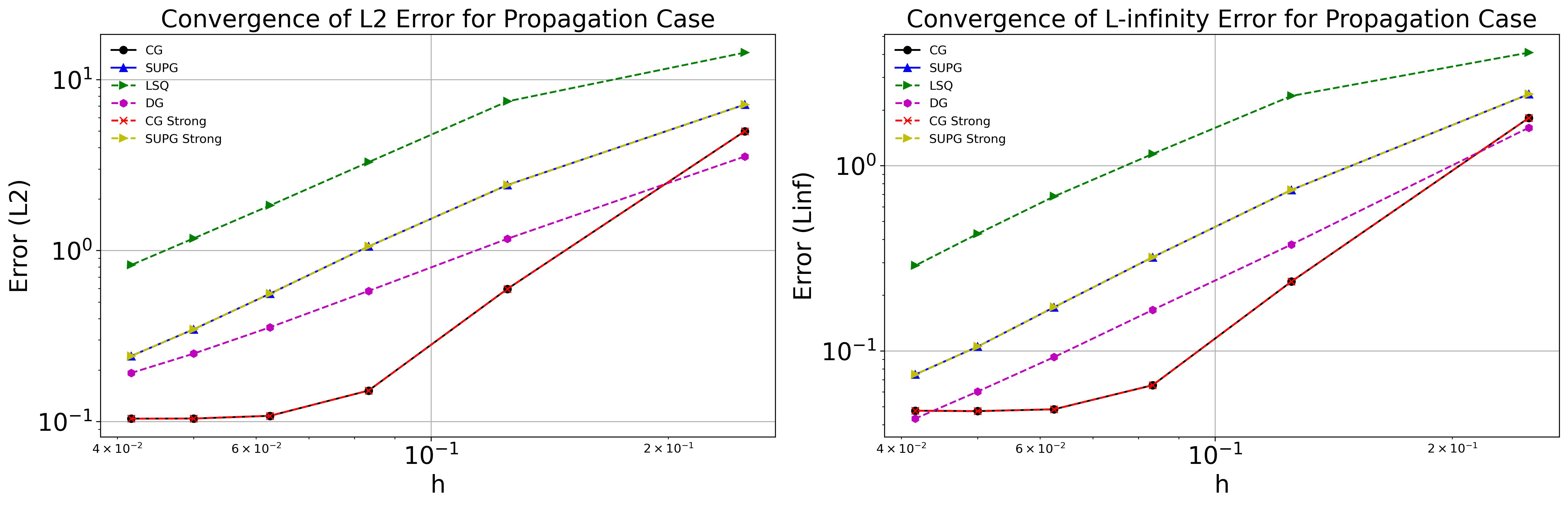}
    \caption{$L^2$ and $l^{\infty}$ errors at T = 5 for all schemes for case 1.}
    \label{fig:PropErrorPlot}
\end{figure}

Each scheme introduced in Section~\ref{chap:stabilized_methods} is discretized with uniform triangular elements in space and order 1 Lagrange finite element functions for both trial and test space. Each scheme is fully implicit in time with order 1 implicit Euler finite difference approximation. The computational grid is refined in space with a uniform spacings of $L/4$, $L/8$, $L/12$, $L/16$, $L/20$, $L/24$. For each finite element scheme, the global $L^2$ and $l^{\infty}$ errors are computed. The errors at the final time $T = 5$  are plotted for both norms and for all schemes in \textbf{Figure~\ref{fig:PropErrorPlot}}. 

For the smooth solution present in case 1, nearly optimal convergence rates of $O(h^{k+1})$ were shown for the SUPG cases. The CG methods without stabilization converged at a rate of $h^2$ until machine precision was met at $L/20$, which brings down the average shown in~\textbf{Table~\ref{tab:2DRates}} to 1.559. 
\begin{table}[h!]
\centering
\caption{\label{tab:2DRates}  Error convergence rates for case 1.}
\begin{tabular}{|l|l|l|l|l|}
\hline
{Method} &   {Avg. L2 rate} & {Avg. Linf rate}  \\
\hline
CG Strong  & 1.559 &  1.445 \\
\hline
CG & 1.559 &  1.445 \\
\hline
SUPG strong & 1.988 & 2.01  \\
\hline
SUPG & 1.988 & 2.01  \\
\hline
Least Squares & 1.790 & 1.730  \\
\hline
DG & 1.611 & 1.981  \\
\hline
\end{tabular}
\end{table}
The errors in the SUPG converged at a better rate than its theoretical rate of $O(h^{1.5})$ most likely due to the minimal amount of residual present in the smooth solution. Least squares had the highest in magnitude of $L^2$ error and converged at a rate between $h^{1.5}$ and $h^{2}$. Lastly, we observe that the DG method had the lowest magnitude in errors for stabilized schemes and converged near the optimal theoretical rate of $h^{1.5}$.

\subsection{Formation of a Shock Near the Boundary}
This test case is an adaptation of a common numerical test for advection dominated PDEs and is adapted from the work of Egger and Sch{\"o}berl~\cite{egger2010hybrid}. This case has an analytic solution that begins as a smooth bump then shifts into a shock towards the right boundary. The aim is to see how the stabilized schemes handle ill-conditioned problems and formation of discontinuities. \markl{Stability in this situation is critical for an operational model because a shock in the solution may form from sharp changes in bathymetry, currents, or wind forcing.}

\begin{figure}[h!]
    \centering
    \includegraphics[width=\textwidth]{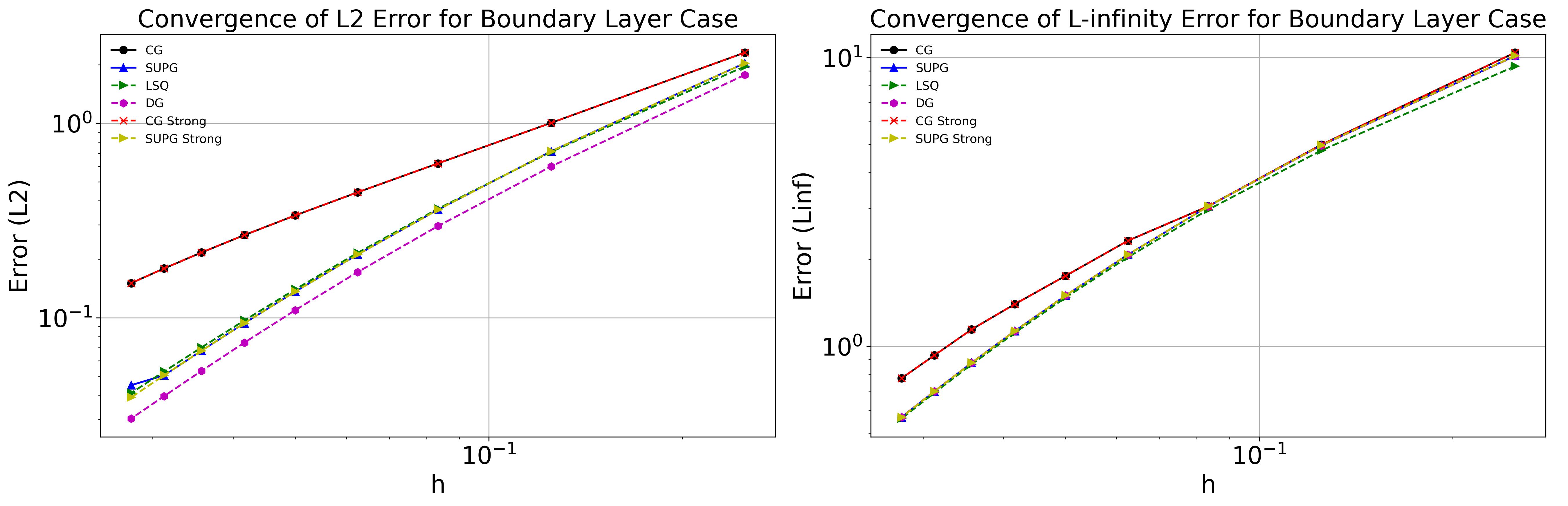}
    \caption{$L^2$ and $l^{\infty}$ errors at T = 1 for all schemes for case 2.}
    \label{fig:Case2Err}
\end{figure}

\begin{figure}[h!]
    \centering
    \includegraphics[width=\textwidth]{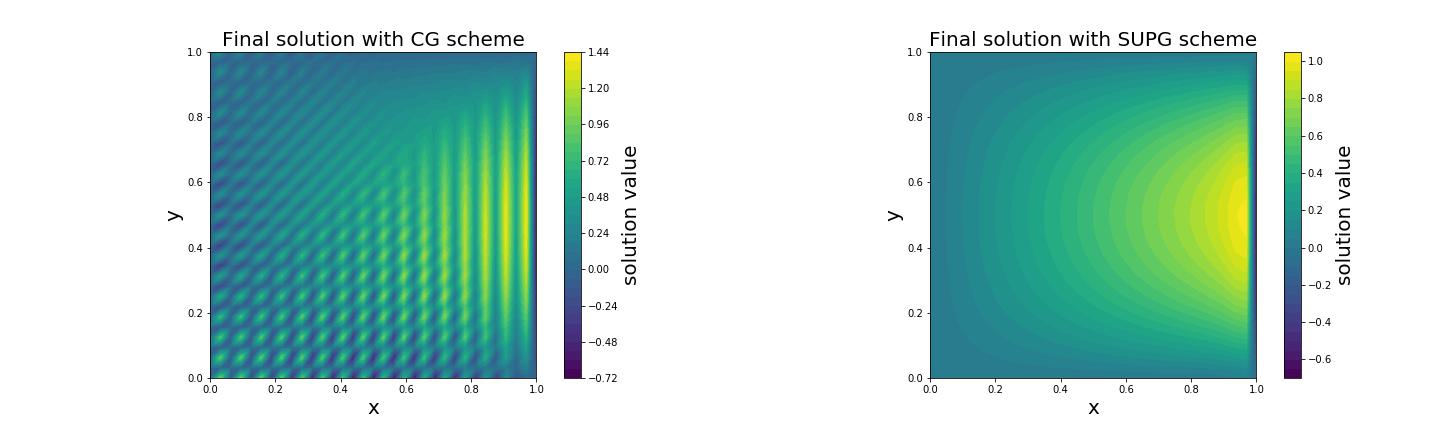}
    \caption{Solution for case 2 on a 32x32 node grid, SUPG right CG left.}
    \label{fig:show_osc}
\end{figure}

The domain is the unit square $\Omega = (0, 1) \times (0,1)$ with a uniform velocity field
$\textbf{c} = \{1,0\}$. The analytic solution is:
\begin{equation}
N_{exact} =  (-4(y-0.5)^2+1)\left[ x + \frac{e^{\gamma \mathbf{c}_x x} - 1 }{1 - e^{\gamma \mathbf{c}_x}}\right],
\end{equation}
where $\gamma = 100t$, $\mathbf{c}_x = 1$, and $t\in[0.01,1]$. The number of time steps for each run is set to 1000. A consequence of this analytic solution is that S should be:
\begin{equation}
    (-4(y-0.5)^2+1)\left[\frac{100 e^{100t}(e^{100tx} - 1)}{(1-e^{100t})^2} + \frac{100xe^{100tx}}{1-e^{100t}}\right] + (-4(y-0.5)^2+1)\left[ 1 + \frac{100t e^{\gamma \cdot \mathbf{c}_x \cdot x} }{1 - e^{\gamma \cdot \mathbf{c}_x}}\right].
\end{equation}
The domains are the same structure as in case 1, and the levels of h refinement for this case are $L/4$, $L/8$, $L/12$, $L/16$, $L/20$, $L/24$, $L/28$, $L/32$, $L/36$. The errors at final time $T=1$ are plotted for both $L^2$ and $l^{\infty}$ norms are displayed in \textbf{Figure \ref{fig:Case2Err}}.
Even with a small time step that is guaranteed stable, the solutions from the CG weak forms both produced the infamous oscillations in the final solution. A comparison in the solution at $T=1$ of CG and SUPG implementations are shown in~\textbf{Figure~\ref{fig:show_osc}}.
\begin{table}[h!]
\centering
\caption{\label{tab:2DRatesB}  Error convergence rates for case 2.}
\begin{tabular}{|l|l|l|l|l|}
\hline
{Method} &  {Avg. L2 rate} & {Avg. Linf rate}  \\
\hline
CG Strong  &   1.287 & 1.270\\
\hline
CG &  1.287 & 1.270\\
\hline
SUPG strong  & 1.948 & 1.461 \\
\hline
SUPG & 1.795 & 1.461 \\
\hline
Least Squares  & 1.907 & 1.440 \\
\hline
DG &  1.999 & 1.460 \\
\hline
\end{tabular}
\end{table}
Consequently, the  resulting error convergence rates in the CG implementations were not optimal as in the former case case. Interestingly, the SUPG and SUPG strong results did have differences as presented in \textbf{Figure~\ref{fig:Case2Err}} with the SUPG strong version performing best. All  the stabilized schemes did not produce visible oscillations at any level of refinement and all schemes converged at equal to or higher than the best theoretical rates as presented in \textbf{Table~\ref{tab:2DRatesB}}. The DG method had the lowest error of all the stabilized schemes, which isn't surprising because for the same polynomial order the DG method has nearly double the degrees of freedom as the continuous implementations. \markl{We now move onto the test cases for the full WAE problem. For the full implementation, SUPG was chosen because it performed well in our preliminary 2D studies, it is most simple to implement and computationally the cheapest. In the future we would like to also implement the Least Squares and DG schemes.}

\section{Full Scale Test Cases}\label{chap:Wavex_vv} 
In this section, 6 test cases from the ONR testbed will be implemented. For each of the cases the RMSE and $l^{\infty}$ errors will be computed with respect to the WAVEx output and analytic solution or lab data depending on the case. The RMSE is computed using the definition:
\begin{equation}\label{eqn:RMSE}
    RMSE(x,y) = \sqrt{\frac{\sum_{i=1}^{N_d}(x_i-y_i)^2}{N_d}},
\end{equation}
where $x$ represents the set of parameters generated by the wave model, $y$ is the set of parameters that come from either the analytic solution or observation, and $N_d$ is the number of points the ONR test bed has data for. Using the same nomenclature as in~\eqref{eqn:RMSE}, the $l^{\infty}$ measure is defined as:
\begin{equation}
    l^{\infty}(x,y) = \max_{1\leq i \leq N_d}\lvert x_i-y_i \rvert.
\end{equation}
In all cases, the data that is compared to the WAVEx model is either significant wave height ($H_s$), or both $H_s$ and mean wave direction ($\theta_{mean}$). Significant wave height is defined as the highest $1/3$ of recorded waves at a point, and mean wave direction is \markl{the mean direction of the spectrum}. In WAVEx, the aforementioned statistical values must be estimated. The significant wave height is estimated by first calculating the zeroth moment of the spectrum, $m_0$. In general the $i^{th}$ moment of the spectrum is defined as:
\begin{equation}\label{eqn:m0}
    m_i(x,y) = \int_{-\pi}^{\pi}\int^{\sigma_{max}}_{\sigma_{min}} \sigma^i E(x,y,\sigma,\theta) d\sigma d\theta,
\end{equation}
and recall $E = N\sigma$. The significant wave height is subsequently estimated by:
\begin{equation}\label{eqn:Hs}
    H_s(x,y) \approx 4 \sqrt{m_0}.
\end{equation}
The reasoning behind this being a decent approximation to the highest one third of waves is not trivial but its assumptions and derivation can be found in several books such as the~\cite{holthuijsen_2007,leblond1981waves}. The mean wave direction in degrees is computed as:
\begin{equation}\label{eqn:meanwave}
    \theta_{mean}(x,y) = \frac{180}{\pi} \frac{\int_{-\pi}^{\pi}\int^{\sigma_{max}}_{\sigma_{min}} \cos{(\theta)}E(x,y,\sigma,\theta) d\sigma d\theta }{\int_{-\pi}^{\pi}\int^{\sigma_{max}}_{\sigma_{min}} sin(\theta) E(x,y,\sigma,\theta) d\sigma d\theta}.
\end{equation}

In all the presented test cases, WAVEx is run using the SUPG strong scheme as in~\eqref{eqn:semi-SUPGI} and the implicit time stepping uses the Crank-Nicolson scheme, which is when $\alpha_t = 0.5$ from~\eqref{eqn:fulldiscrete} for second order accuracy in time. 


%
%
%
%

\subsection{A21: Shoaling}

\begin{figure}[h!]
\begin{center}
    \centering
    \includegraphics[width=\textwidth]{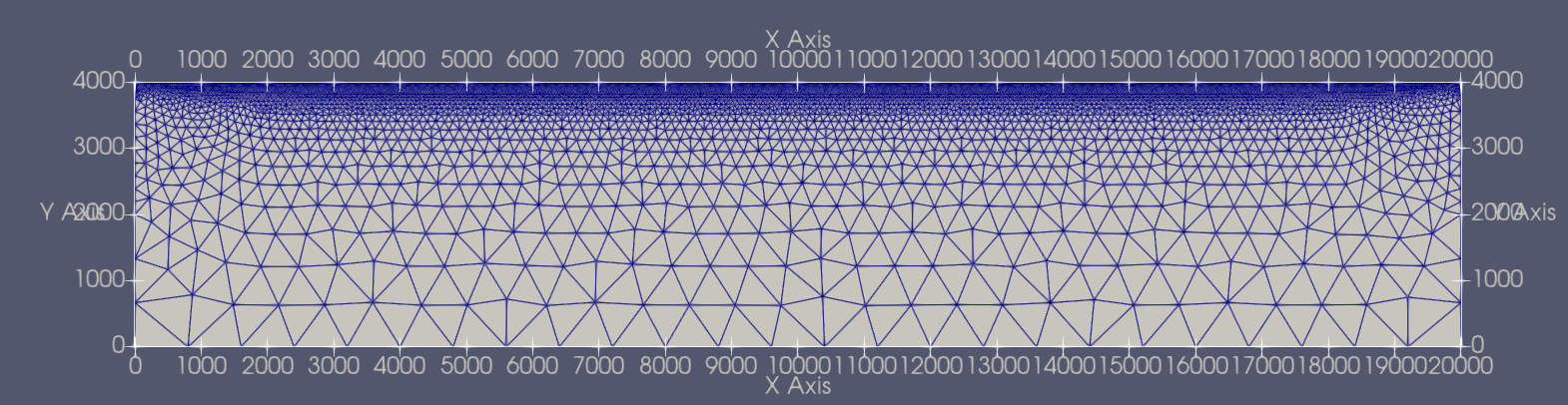}
    \caption{The unstructured geographic mesh used for test cases A11, A21.}
    \label{fig:Unstructured_grid}
\end{center}
\end{figure}

\begin{figure}[h!]
\begin{center}
    \centering
    \includegraphics[width=\textwidth]{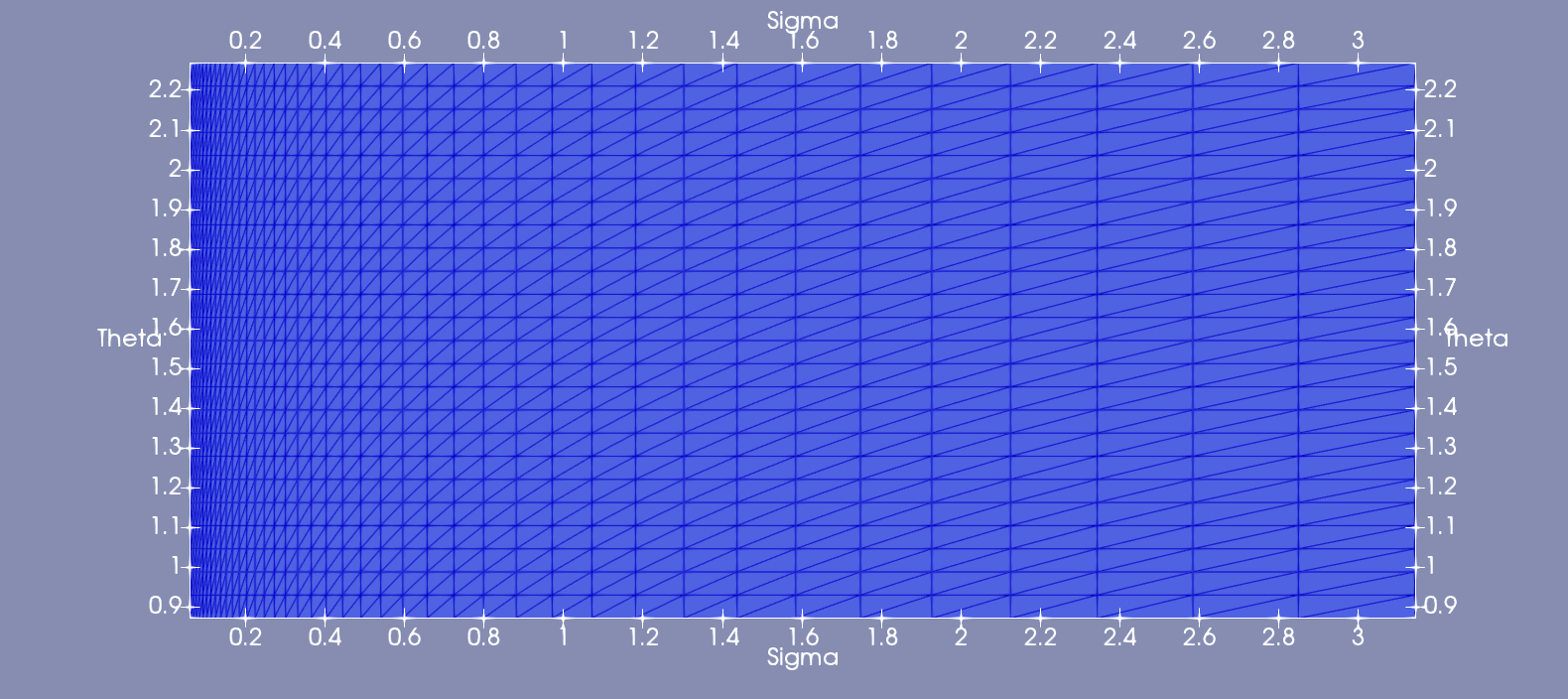}
    \caption{Logarithmic spaced spectral grid as used in cases A11, A21.}
    \label{fig:Spectral_grid}
\end{center}
\end{figure}

\begin{figure}[h!]
\begin{center}
    \centering
    \includegraphics[width=\textwidth]{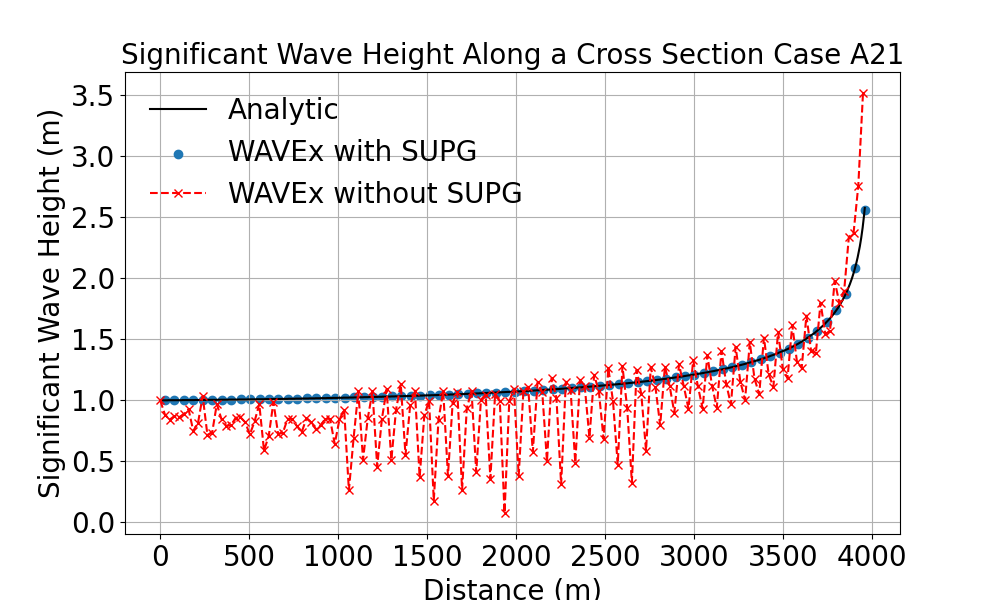}
    \caption{Significant wave heights from case A21.}
    \label{fig:A21_HS}
\end{center}
\end{figure}

The first test case is an idealized situation where some waves are propagating from the deep ocean directly into the coastline. The water depth in the domain is uniformly decreasing from 20 meters to 0 meters with a slope of 0.005. The waves are propagating perpendicular to the coastline and are approximately monochromatic. The waves at the inflow boundary have a significant wave height of 1 meter with a mean period of 10 seconds. All source terms are zero and there are no currents in this case. Because all source terms are zero, the analytic solution for the significant wave height can easily be computed because the WAE becomes a pure advection equation with nonuniform velocity field. The result is that the significant wave height should be:
 \begin{equation}\label{eq:A21solution}
     H_s(x) =  H_s(0)\sqrt{\frac{c_g(0)}{c_g(x)}}.
 \end{equation}
An unstructured mesh is used in the WAVEx simulations for cases A11 and A21 illustrated in~\textbf{Figure~\ref{fig:Unstructured_grid}}. The mesh has 7458 nodes comprising 13841 triangular elements and has variable resolution with element size ranging from 800 meters near on the bottom to 20 meters near the top.
In WAVEx, the spectral grid is capable of being unstructured but for all of the ONR cases it is structured with logarithmic spacing in frequency and uniform in direction. An example is shown in \textbf{Figure~\ref{fig:Spectral_grid}}.
The time step is set to 50 seconds for all the cases, and WAVEx solves until the iterative solver reaches a steady state. The significant wave height is computed along the propagation direction and results are plotted in~\textbf{Figure~\ref{fig:A21_HS}}. The results in this figure also clearly demonstrate the effect of instability in the Bubnov-Galerkin solution as its oscillations makes the solution very poor.
The computed RMSE error vs the analytic solution is 0.000695 m for WAVEx with SUPG stabilization and 0.313 m without, while $l^{\infty}$ error is 0.0018 m with SUPG and .9916 m without. 
\subsection{A11: Refraction}
\begin{figure}[h!]
\begin{center}
    \centering
    \includegraphics[width=\textwidth]{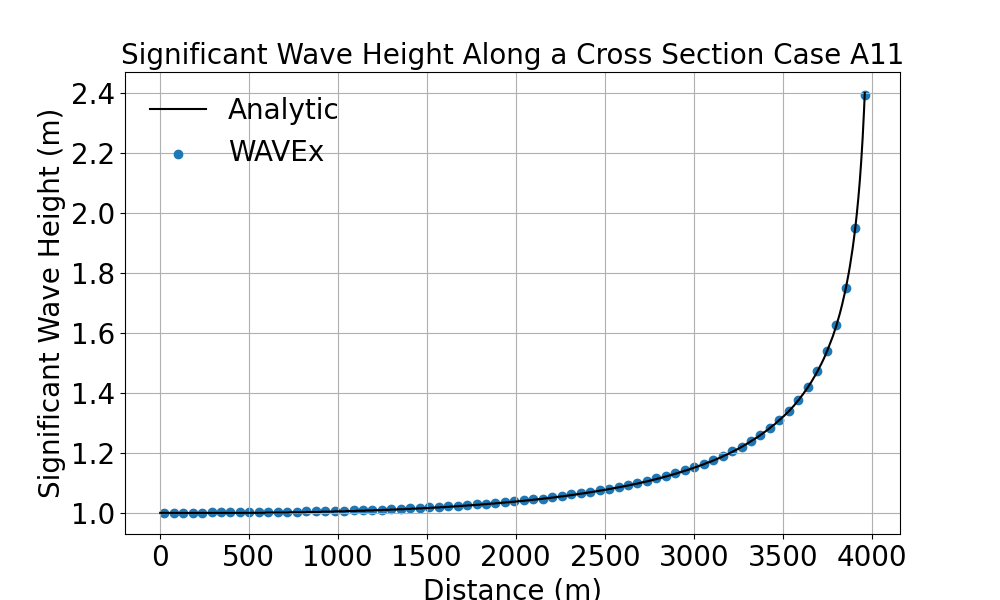}
    \caption{Significant wave height for case A11.}
    \label{fig:A11_HS}
\end{center}
\end{figure}

\begin{figure}[h!]
\begin{center}
    \centering
    \includegraphics[width=\textwidth]{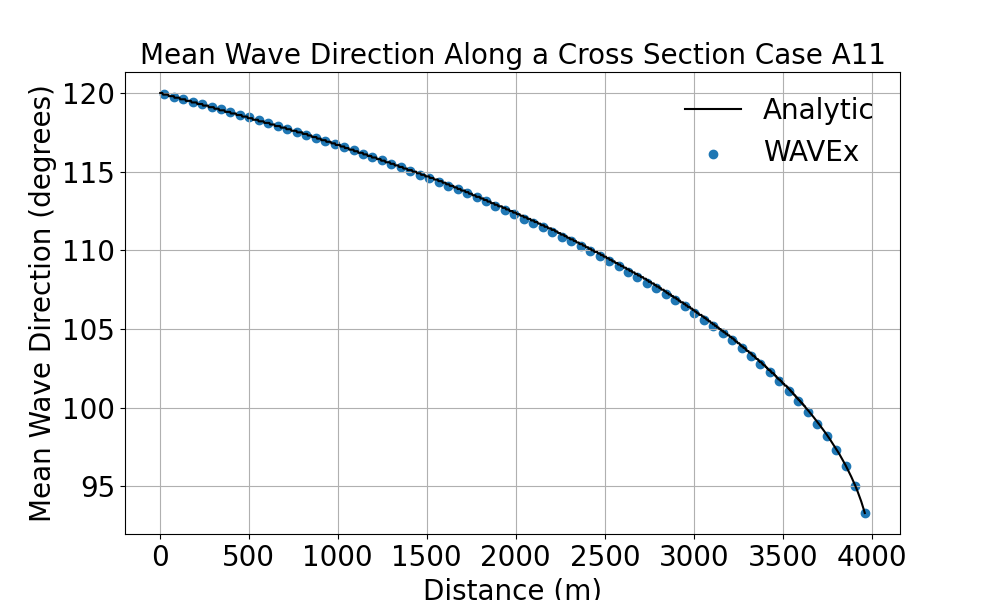}
    \caption{Mean wave direction for case A11.}
    \label{fig:A11_DIR}
\end{center}
\end{figure}

The second test case is similar to case A21 in that the bathymetry is the same and the domain is identical. The only difference is that the waves on the inflow boundary now approach the shoreline at a direction 30 degrees above the direction perpendicular to the shoreline.  This too has an analytic solution. Again, no source terms are present and the currents are all 0. In addition to significant wave height, the mean wave direction is measured and the analytic solutions are:
\begin{equation}\label{eqn:A11Solution}
\begin{split}
        \theta(x) = arcsin(c(x)\frac{sin(30)}{c(0)}),\\
        H_s(x) =  H_s(0)\sqrt{\frac{c_g(0) cos(30)}{c_g(x)cos(\theta(x))}}.
\end{split}
\end{equation}
The significant wave height is plotted against the analytic solution in~\textbf{Figure~\ref{fig:A11_HS}} while the mean wave direction is plotted in~\textbf{Figure~\ref{fig:A11_DIR}}. In both figures, the WAVEx and analytic results agree very well. The observed RMSE error in $H_s$ is 0.00261 m and RMSE error in degrees is 0.119 degrees while the $l^{\infty}$ error is 0.00865 m and 0.1946 degrees.
\subsection{A31-34: Currents}
Cases A31 through A34 are situations in deep water where there are interacting currents but no source terms. All simulations are performed on the same domain, and is a rectangle shape of constant arbitrarily deep water (set to 10000m) and is 4000 meters long in the principal direction. To avoid any noise from the side boundaries, the width of the domain is set to 10000 meters. Each case has either different currents or different mean wave directions on the inflow boundary. Case A31 has an inflow boundary where the mean wave direction is parallel with the principle axis with a current that is also parallel with the principle axis and increases from 0 m/s at the inflow boundary to 2 m/s at the outflow boundary. Case A32 has the same boundary conditions as A31 but the currents in this case decrease from 0 m/s at the inflow boundary to -2 m/s at the outflow boundary. Case A33 has the same current profile as case A31 but the inflow boundary condition now has a mean wave direction of 120 degrees relative to the positive direction of the principle axis. Finally, case A34 is the same as A33 but instead of 120 degrees, it is set to 60 degrees. All waves at the inflow boundary have a significant wave height of 1 meter and peak period of 10 seconds. Similar to cases A11 and A21, an analytic expression for the directions and significant wave heights may be derived. These are  more complex due to the nonzero currents and are different for cases A31, A32 and the A33, A34 pair. The solutions are different because the former have incoming spectra parallel to the currents while the latter are at an angle. The analytic solution for significant wave height in cases A31 and A32 is:
\begin{equation}\label{eqn:A31-A32solution}
    \begin{split}
        H_s(x) = H_s(0) \sqrt{\frac{c(0)^2}{c(x)(c(x)+2u(x))}},\\
        c(x) = c(0)(0.5 + 0.5\sqrt{1+4\frac{u(x)}{c(0)}}),
    \end{split}
\end{equation}
while the solution for cases A33 and A34 along with the analytic solution for mean wave direction is:
\begin{equation}
    \begin{split}
        H_s(x) = H_s(0)\sqrt{\frac{sin(2\theta(0)}{sin(2\theta(x))}},\\
        \theta(x) = arccos(\frac{gk(0)cos(\theta(0))}{(\frac{2\pi}{Pkper} - u(x) k(0) cos(\theta(0)))^2}).
    \end{split}
\end{equation}

\begin{figure}[h!]
\begin{center}
    \centering
    \includegraphics[width=\textwidth]{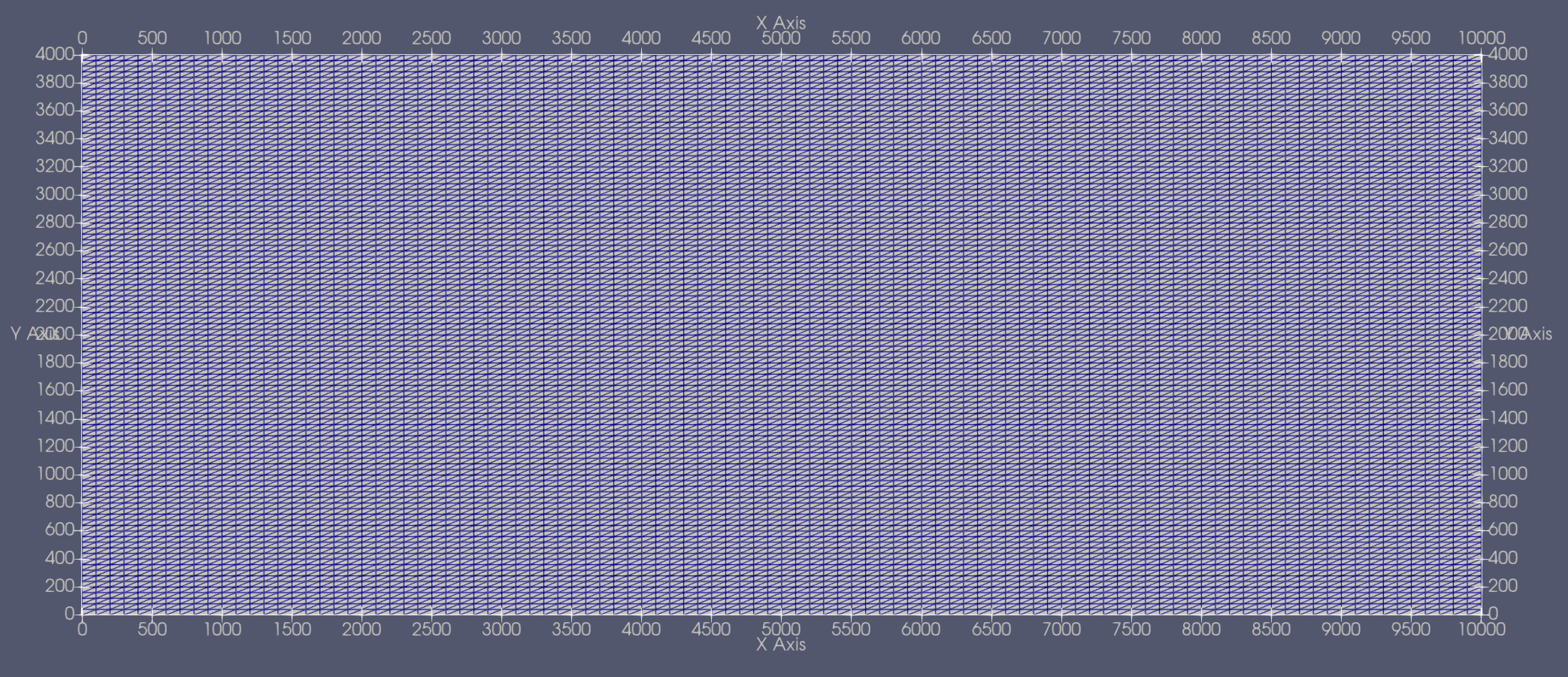}
    \caption{Geographic mesh for cases A31-A34.}
    \label{fig:Currents_grid}
\end{center}
\end{figure}

\begin{figure}[h!]
\begin{center}
    \centering
    \includegraphics[width=\textwidth]{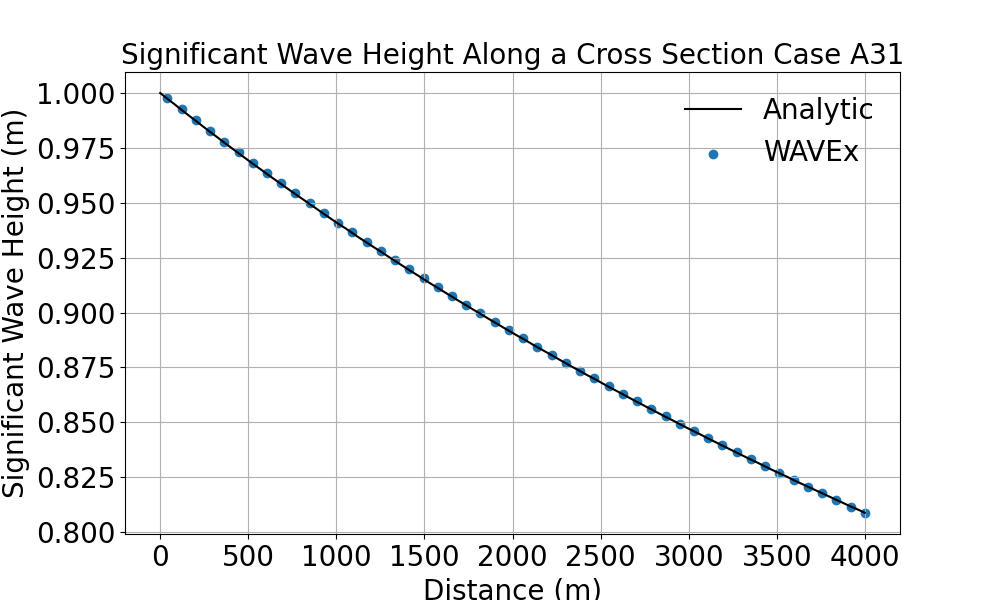}
    \caption{Significant wave height from case A31.}
    \label{fig:A31_HS}
\end{center}
\end{figure}

\begin{figure}[h!]
\begin{center}
    \centering
    \includegraphics[width=\textwidth]{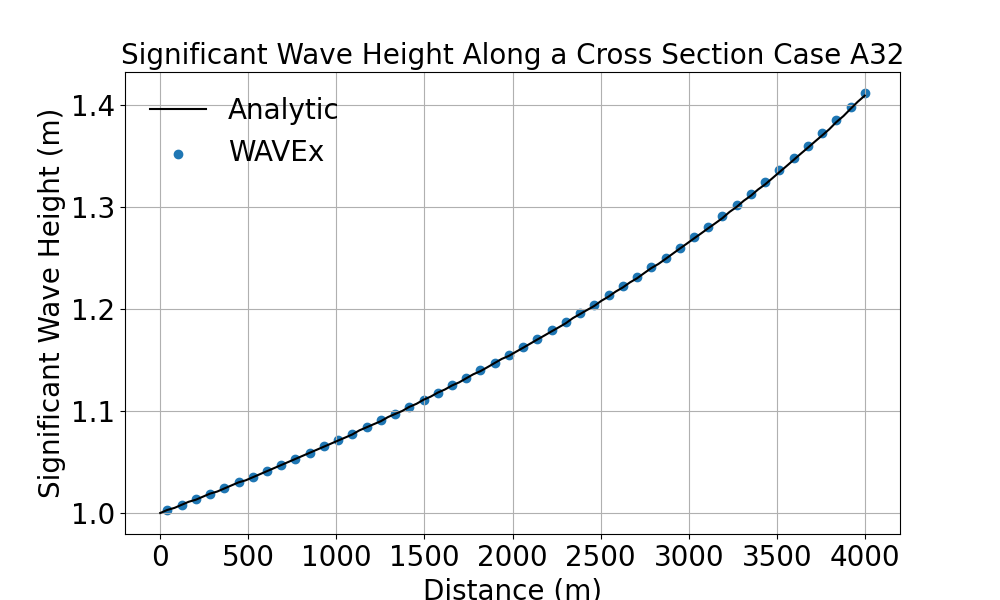}
    \caption{Significant wave height from case A32.}
    \label{fig:A32_HS}
\end{center}
\end{figure}

\begin{figure}[h!]
\begin{center}
    \centering
    \includegraphics[width=\textwidth]{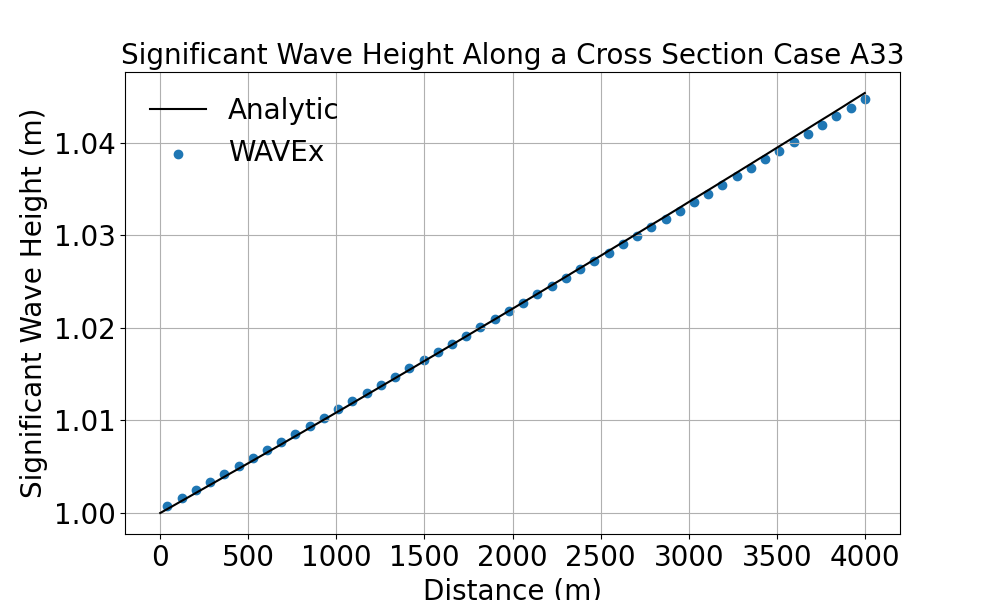}
    \caption{Significant wave height from case A33.}
    \label{fig:A33_HS}
\end{center}
\end{figure}

\begin{figure}[h!]
\begin{center}
    \centering
    \includegraphics[width=\textwidth]{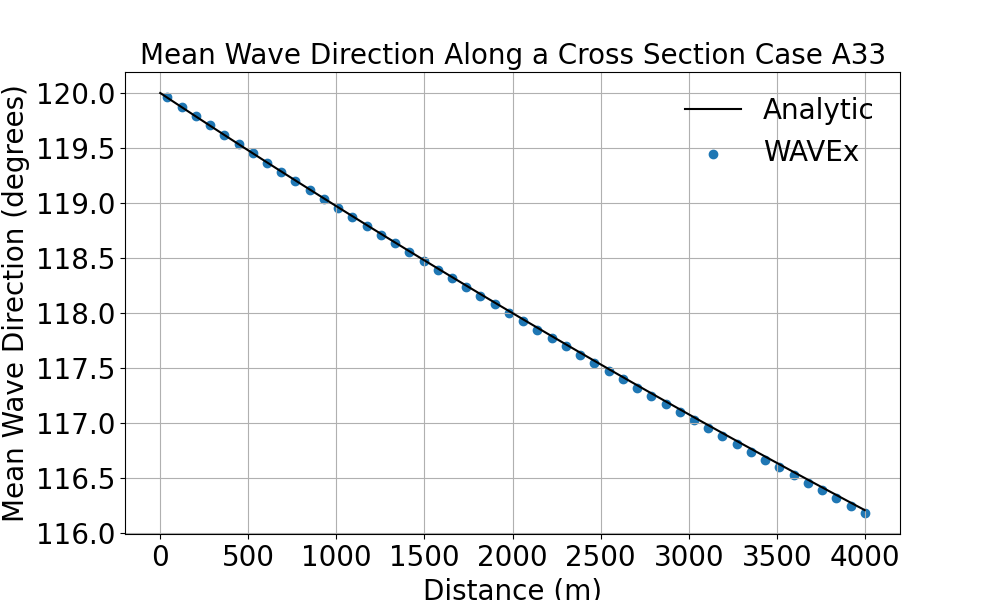}
    \caption{Mean wave direction from case A33.}
    \label{fig:A33_DIR}
\end{center}
\end{figure}

\begin{figure}[h!]
\begin{center}
    \centering
    \includegraphics[width=\textwidth]{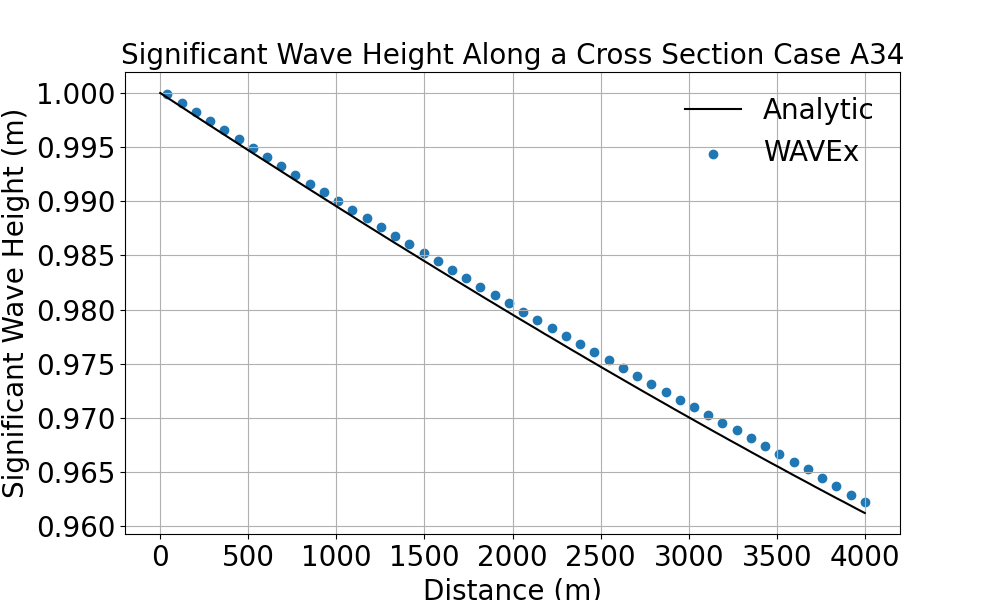}
    \caption{Significant wave height from case A34.}
    \label{fig:A34_HS}
\end{center}
\end{figure}

\begin{figure}[h!]
\begin{center}
    \centering
    \includegraphics[width=\textwidth]{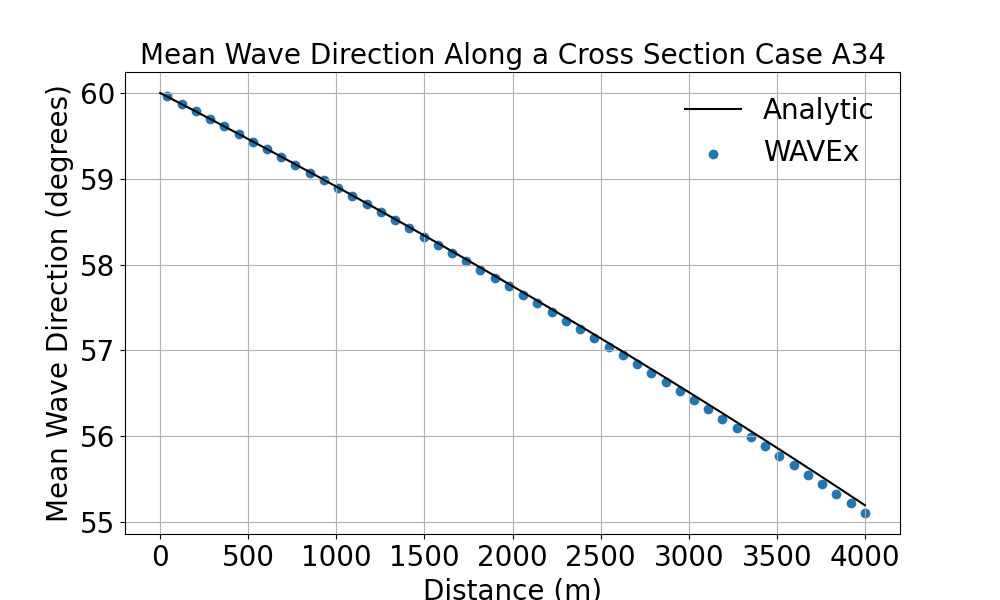}
    \caption{Mean wave direction from case A34.}
    \label{fig:A34_DIR}
\end{center}
\end{figure}

For these cases, the mesh is uniformly spaced with triangular elements. The mesh contains 10,201 nodes that comprise 20,000 elements. This mesh is 10000 meters in the x direction and 4000 meters in the y direction. Each element has a size of around 110 meters. The mesh is 100 elements wide and 100 elements tall. The mesh is illustrated in \textbf{Figure}~\ref{fig:Currents_grid}.
The resulting significant wave heights from case A31 and A32 are shown in~\textbf{Figures~\ref{fig:A31_HS},\ref{fig:A32_HS}}, and the computed RMSE are 0.000255 m, and 0.00109 m respectively while $l^{\infty}$ error is 0.0003737 m and 0.00257 m. Significant wave height for cases A33 and A34 are plotted in~\textbf{Figure~\ref{fig:A33_HS}} and~\textbf{Figure~\ref{fig:A34_HS}}. The RMSE for significant wave height for A33 is 0.000315 m, while A34 is 0.000899 m. The $l^{\infty}$ for case A33 is 0.000659 m while A34 is 0.00128 m. The resulting mean wave direction for cases A33 and A34 are shown in~\textbf{Figure~\ref{fig:A33_DIR}} and~\textbf{Figure~\ref{fig:A34_DIR}}. The RMSE for mean wave direction in case A33 is 0.1793 degrees while for case A34 it is 0.436. The $l^{\infty}$ for case A33 is 0.028 degrees while case A34 is 0.0436 degrees.  We observe close agreement of WAVEx and the analytic solutions in all cases. This agreement is further witnessed in the low errors in both mean wave direction and significant wave height.
\subsection{L11: Wave Breaking on a Beach}

\begin{figure}[h!]
\begin{center}
    \centering
    \includegraphics[width=\textwidth]{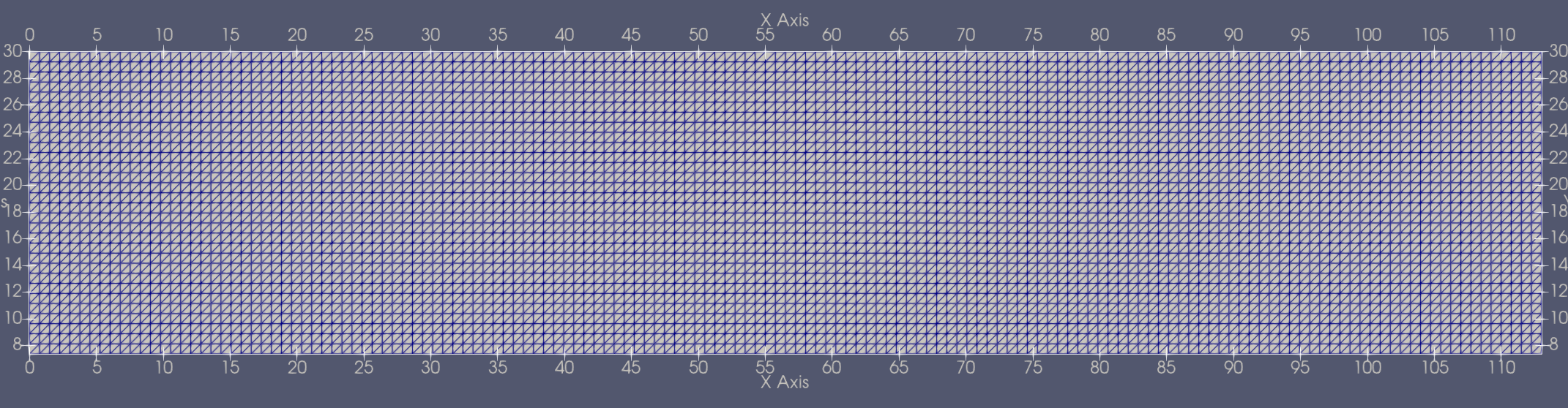}
    \caption{Geographic mesh for case L11.}
    \label{fig:L11_grid}
\end{center}
\end{figure}

\begin{figure}[h!]
\begin{center}
    \centering
    \includegraphics[width=\textwidth]{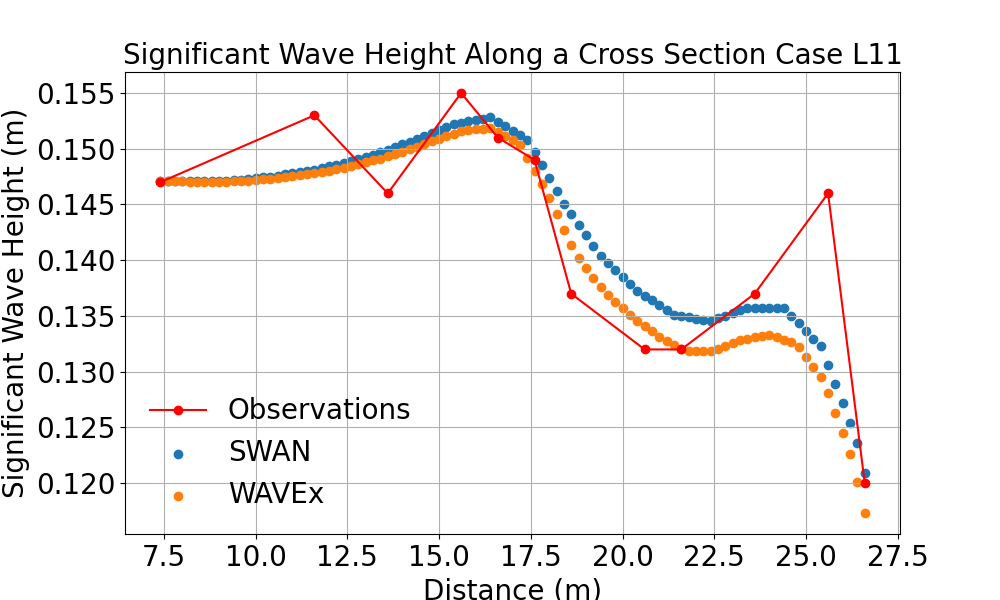}
    \caption{Significant wave height for case L11.}
    \label{fig:L11_HS}
\end{center}
\end{figure}

The final test case is the L11 case, a laboratory experiment intended to simulate groups of waves breaking on a beach. The bathymetry varies in slope as it approaches the beach but is uniform in the along-shore direction. The input spectrum is a JONSWAP spectrum with peakedness parameters specified in the manual~\cite{ONR_testbed}. 
The  mesh that is used is in this case and is set as the rectangle $[0,113] \times [7.4,30]$, it is a structured, uniform mesh with triangular elements. Each element is approximately 1 meter wide and there are 4681 total nodes comprising 9000 elements. The mesh is shown in Figure~\ref{fig:L11_grid}.
In addition to comparison to the measured results from the ONR testbed~\cite{ONR_testbed}, we also compare the WAVEx results to results from SWAN. The SWAN numerical model is set up with the third generation source terms are turned on, however there is no wind in this case. The significant wave heights are recorded both from SWAN and WAVEx and plotted against the observations in~\textbf{Figure~\ref{fig:L11_HS}}. The RMSE in significant wave height with respect to the observations were 0.0064m for WAVEx and 0.00556m for SWAN while the $l^{\infty}$ errors were 0.0179m and 0.015m respectively. \markl{We note that SWAN was run in 1-D mode while WAVEx was used in full 2-D mode, we suspect this could have a slight impact in the results. }

\section{Conclusions}\label{chap:Concl}
The purpose of this work was to develop a finite element spectral wind wave model using the open source FEniCSx library, which we call WAVEx. The motivation behind this was to create a platform so that different numerical schemes and source terms can easily be implemented and tested in the future. First, some stabilized FEM methods were introduced (least squares, SUPG, DG) and convergence rates with respect to mesh refinement were verified in simple 2D test cases. All of the stabilized schemes demonstrated convergence in the $L^2$ and $l^{\infty}$ norms with respect to $h$ refinement while also reducing oscillations when compared to the standard Bubnov-Galerkin CG approach that produced oscillations when the solution became less smooth.

Next, a spectral wave model with SUPG stabilization in 4 dimensions was implemented. The WAE was split into propagation and source terms via Strang splitting for computational efficiency. The propagation portion was implemented with an implicit time stepping scheme while the (nonlinear) source term portion was implemented with an explicit time stepping scheme in order to keep the discrete problem linear. The new wave model showed accuracy and stability in the analytic test cases and a laboratory test case that included source terms.

In the future, a first step to improving WAVEx would be to introduce a custom FEniCSx assembler in order to speed up run times. One drawback of the way WAVEx is currently implemented is that the assembly routine is relatively slow, but this can be circumvented by developing new tools in the FEniCSx. Other future investigations include how WAVEx scales with more complex test cases in addition to fully implicit time stepping by extending WAVEx to solve nonlinear problems. Additionally, implementing the DG and least squares schemes in the full (4D) operational setting is a natural extension that we have not done at the present time. While we have considered several FEMs, there is a plethora of methods of interest that show promise in handling problems like the WAE such as Lagrange-Galerkin or discontinuous Petrov-Galerkin (DPG)~\cite{DEMKOWICZ20101558} methods. 

\section{Acknowledgements}

Author Loveland has been supported by the CSEM Fellowship from the Oden Institute at the University of Texas at Austin. Authors Loveland, Valseth, and Dawson have been supported by the United States National Science Foundation - NSF PREEVENTS Track
2 Program, under NSF Grant Number 1855047 and the Department of Homeland Security Coastal Resilience Center research project "Accurate and Fast Wave Modeling and Coupling with ADCIRC".
Author Meixner has been supported by the National Oceanic and Atmospheric Administration. The authors would also like to thank the reviewers of this manuscript for their time and thoughtful suggestions.

\appendix

\section{Appendix A}\label{ap:B}
 This appendix illustrates how to rewrite both~\eqref{eqn:semi-SUPGI} and ~\eqref{eqn:semi-SUPGII} into a product of 2-dimensional integrals so that FEniCSx can interpret it through the unified form language (UFL) framework.
We evaluate the third term in~\eqref{eqn:semi-SUPGII}:
\begin{equation}
\begin{split}
        (\mathbf{c}\phi_i\psi_j \cdot \mathbf{n},\gamma_k\beta_l)_{\partial \Omega} = (\mathbf{c}\phi_i\psi_j \cdot \mathbf{n},\gamma_k\beta_l)_{\partial \Omega_1 \times \Omega_2 \cup \Omega_1 \times \partial \Omega_2} =  \\ \int_{\Omega_2} \psi_j \beta_l \int_{\partial \Omega_1} \phi_i \gamma_k\mathbf{c_1} \cdot \mathbf{n}_1 dxdy + \int_{\partial \Omega_2} \psi_j \beta_l \mathbf{n}_2 \cdot \int_{\Omega_1} \mathbf{c_2} \phi_i \beta_k dxdy,
\end{split}
\end{equation}
and for the continuous Galerkin components of the form in~\eqref{eqn:semi-SUPGI} we have:
\begin{equation}
    (\frac{\partial N}{\partial t} + \nabla \cdot (\mathbf{c}N) - S, v)_{\Omega}. 
\end{equation}
Replacing the functions with the finite element basis and omitting the transient and source term we get:
\begin{equation}
\begin{split}
    (\nabla \cdot (\mathbf{c} \phi_i \psi_j) ,  \gamma_k \beta_l)_{\Omega} = (\mathbf{c} \cdot \nabla(\phi_i\psi_j) + (\phi_i\psi_j) \nabla \cdot \mathbf{c} , \gamma_k\beta_l)_{\Omega} =,\\
    (\mathbf{c_1} \cdot \psi_j\nabla_1\phi_i + \mathbf{c_2} \cdot \phi_i \nabla_2 \psi_j + \phi_i\psi_j \nabla_1 \cdot \mathbf{c_1} + \phi_i \psi_j \nabla_2 \cdot \mathbf{c_2}, \gamma_k\beta_l)_{\Omega},
\end{split}
\end{equation}
which decomposes into the following terms:
\begin{equation}
    \begin{split}
        (\mathbf{c_1} \cdot \psi_j\nabla_1\phi_i, \gamma_k\beta_l)_{\Omega} = \int_{\Omega_2} \psi_j\beta_l \int_{\Omega_1} \mathbf{c_1} \cdot \nabla_1 (\phi_i) \gamma_k dxdy, \ \\
        (\mathbf{c_2} \cdot \phi_i \nabla_2 \psi_j , \gamma_k\beta_l)_{\Omega} = \int_{\Omega_2} \beta_l \nabla_2 \psi_j \cdot \int_{\Omega_1} \mathbf{c_2} \phi_i \gamma_k dxdy,\\
        (\phi_i\psi_j \nabla_1 \cdot \mathbf{c_1}, \gamma_k\beta_l)_{\Omega} = \int_{\Omega_2} \psi_j \beta_l \int_{\Omega_1} \nabla_1 \cdot (\mathbf{c_1}) \phi_i \gamma_k dxdy,\\
        (\phi_i \psi_j \nabla_2 \cdot \mathbf{c_2}, \gamma_k\beta_l)_{\Omega} = \int_{\Omega_1}\phi_i \gamma_k \int_{\Omega_2} \nabla_2 \cdot (\mathbf{c_2}) \psi_j \beta_l dydx.
    \end{split}
\end{equation}

Now for SUPG stabilization terms in both forms we need to add the upwind component:
\begin{equation}
    (\frac{\partial N}{\partial t} + \nabla \cdot (\mathbf{c}N) - S, \tau \mathbf{c} \cdot \nabla (v))_{\Omega}. 
\end{equation}
The source term will be absorbed into the RHS and we will first focus our attention on the streamline diffusion term:
\begin{equation}
    (\nabla \cdot (\mathbf{c}N), \tau \mathbf{c} \cdot \nabla v)_{\Omega}.
\end{equation}
Using our product basis expansion, we consider this term now in the discrete setting and expand using the product rule:
\begin{equation}
    \begin{split}
    (\nabla \cdot (\mathbf{c}N), \tau \mathbf{c} \cdot \nabla v)_{\Omega} \rightarrow (\nabla \cdot (\mathbf{c} \phi_i\psi_j), \tau \mathbf{c} \cdot \nabla(\gamma_k\beta_l))_{\Omega} = ,\\
    (\nabla \cdot (\mathbf{c} \phi_i\psi_j), \tau \beta_l\mathbf{c}_1 \cdot \nabla_1\gamma_k + \tau \gamma_k \mathbf{c}_2 \cdot \nabla_2\beta_l)_{\Omega} =, \\
    (\mathbf{c} \cdot \nabla(\phi_i\psi_j) + (\phi_i\psi_j) \nabla \cdot \mathbf{c}, \tau \beta_l\mathbf{c}_1 \cdot \nabla_1\gamma_k + \tau \gamma_k \mathbf{c}_2 \cdot \nabla_2\beta_l)_{\Omega} =,\\
    (\mathbf{c_1} \cdot \psi_j\nabla_1(\phi_i) + \mathbf{c_2} \cdot \phi_i\nabla_2(\psi_j) + (\phi_i\psi_j) \nabla_1 \cdot \mathbf{c_1} + (\phi_i\psi_j) \nabla_2 \cdot \mathbf{c_2}, \tau \beta_l\mathbf{c}_1 \cdot \nabla_1\gamma_k + \tau \gamma_k \mathbf{c}_2 \cdot \nabla_2\beta_l)_{\Omega}.
    \end{split}
\end{equation}
We have 4 scalar terms on the left hand part of the inner product and 2 on the right, if we expand all of these out it will lead to a sum of 8 total terms:
\begin{equation}
    \begin{split}
        (\mathbf{c_1} \cdot \psi_j \nabla_1(\phi_i), \tau \mathbf{c_1} \cdot \beta_l \nabla_1 (\gamma_k))_{\Omega},\\
        (\mathbf{c_1} \cdot \psi_j \nabla_1(\phi_i), \tau \mathbf{c_2} \cdot  \gamma_k \nabla_2 (\beta_l))_{\Omega,}\\
        (\mathbf{c_2} \cdot \phi_i \nabla_2(\psi_j), \tau \mathbf{c_1} \cdot \beta_l \nabla_1 (\gamma_k))_{\Omega},\\
        (\mathbf{c_2} \cdot \phi_i \nabla_2(\psi_j), \tau \mathbf{c_2} \cdot  \gamma_k \nabla_2 (\beta_l))_{\Omega},\\
        ((\phi_i\psi_j) \nabla_1 \cdot \mathbf{c_1}, \tau \mathbf{c_1} \cdot \beta_l  \nabla_1 (\gamma_k))_{\Omega},\\
        ((\phi_i\psi_j) \nabla_1 \cdot \mathbf{c_1}, \tau \mathbf{c_2} \cdot  \gamma_k \nabla_2 (\beta_l))_{\Omega},\\
        ((\phi_i\psi_j) \nabla_2 \cdot \mathbf{c_2}, \tau \mathbf{c_1} \cdot  \beta_l \nabla_1 (\gamma_k))_{\Omega},\\
        ((\phi_i\psi_j) \nabla_2 \cdot \mathbf{c_2}, \tau \mathbf{c_2} \cdot  \gamma_k \nabla_2 (\beta_l))_{\Omega}.\\
    \end{split}
\end{equation}
Now looking term by term we have first:
\begin{equation}
    (\mathbf{c_1} \psi_j \cdot \nabla_1 \phi_i, \tau \beta_l \mathbf{c_1} \cdot \nabla_1\gamma_k)_{\Omega} = \int_{\Omega_2} \psi_j \beta_l \int_{\Omega_1} \tau \mathbf{c_1} \cdot \nabla_1 \phi_i  \mathbf{c_1} \cdot \nabla_1 \gamma_k dxdy.
\end{equation}
The second term we get:
\begin{equation}
    (\mathbf{c}_1 \psi_j \cdot \nabla_1 \phi_i,\tau \gamma_k \mathbf{c}_2 \cdot  \nabla_2\beta_l)_{\Omega} = \int_{\Omega_2} \psi_j \nabla_2\beta_l \cdot \int_{\Omega_1} \mathbf{c_2} \tau \gamma_k \mathbf{c_1} \cdot \nabla_1 \phi_i dxdy.
\end{equation}
The third term will be:
\begin{equation}
    (\mathbf{c}_2 \phi_i \cdot \nabla_2 \psi_j , \tau \beta_l \mathbf{c_1} \cdot \nabla_1 \gamma_k)_{\Omega} =
    \int_{\Omega_2} \nabla_2 \psi_j \beta_l \cdot \int_{\Omega_1} \tau \mathbf{c}_2 \phi_i   \mathbf{c_1} \cdot \nabla_1 \gamma_k dx dy.
\end{equation}
Fourth term:
\begin{equation}
    (\mathbf{c_2} \phi_i \cdot \nabla_2 \psi_j, \tau \gamma_k \mathbf{c_2} \cdot \nabla_2 \beta_l)_{\Omega} = 
    \int_{\Omega_1} \phi_i \gamma_k \int_{\Omega_2} \tau \mathbf{c_2} \cdot \nabla_2 \psi_j \mathbf{c_2} \cdot \nabla_2\beta_l dydx.
\end{equation}
Fifth term:
\begin{equation}
    ((\phi_i\psi_j) \nabla_1 \cdot \mathbf{c_1}, \tau \mathbf{c_1} \cdot \beta_l  \nabla_1 (\gamma_k))_{\Omega}
    = \int_{\Omega_2} \psi_j \beta_l \int_{\Omega_1} \tau \phi_i \nabla_1 \cdot \mathbf{c_1} \mathbf{c_1} \cdot \nabla_1(\gamma_k)dxdy .
\end{equation}
Sixth:
\begin{equation}
    ((\phi_i\psi_j) \nabla_1 \cdot \mathbf{c_1}, \tau \mathbf{c_2} \cdot  \gamma_k \nabla_2 (\beta_l))_{\Omega}
    = \int_{\Omega_2} \psi_j \nabla_2(\beta_l) \cdot \int_{\Omega_1} \mathbf{c_2} \tau \phi_i \gamma_k \nabla_1 \cdot \mathbf{c_1} dxdy .
\end{equation}
Seventh:
\begin{equation}
     ((\phi_i\psi_j) \nabla_2 \cdot \mathbf{c_2}, \tau \mathbf{c_1} \cdot \beta_l  \nabla_1 (\gamma_k))_{\Omega} = \int_{\Omega_1} \phi_i \nabla_1(\gamma_k) \cdot \int_{\Omega_2} \mathbf{c_1} \tau \psi_j\beta_l \nabla_2 \cdot \mathbf{c_2} dydx.
\end{equation}
Eighth:
\begin{equation}
    ((\phi_i\psi_j) \nabla_2 \cdot \mathbf{c_2}, \tau \mathbf{c_2} \cdot  \gamma_k \nabla_2 (\beta_l))_{\Omega} = \int_{\Omega_1} \phi_i \gamma_k \int_{\Omega_2} \tau \mathbf{c_2} \cdot \nabla_2(\beta_l)  \psi_j \nabla_2 \cdot \mathbf{c_2} dydx.
\end{equation}
Now the last part that remains is the upwinding portion based on the choice of time step. Choosing a simple implicit time step we get:
\begin{equation}
    (\frac{\partial N}{\partial t}, \tau \mathbf{c} \cdot \nabla v)_{\Omega} \approx
    (\frac{\phi_i^{n+1}\psi_j^{n+1} - \phi_i^n\psi_j^n}{\Delta t}, \tau \mathbf{c} \cdot \nabla (\gamma_k\beta_l))_{\Omega}.
\end{equation}
The component with the previous time step $n$ will go to RHS and the part that is in current time step $n+1$ will be in stiffness matrix. Both descritizations however will be the same so I will just do the $n+1$ time step after multiplying by $\Delta t$:
\begin{equation}
    (\phi_i \psi_j, \tau \mathbf{c} \cdot \nabla(\gamma_k\beta_l))_{\Omega} =  (\phi_i \psi_j, \tau( \mathbf{c}_x \cdot \beta_l \nabla_x(\gamma_k) + \mathbf{c}_y \cdot \gamma_k \nabla_y(\beta_l)))_{\Omega}.
\end{equation}
Focusing on left term we can evaluate using this decomposition:
\begin{equation}
    (\phi_i \psi_j, \tau \mathbf{c}_x \cdot \beta_l \nabla_x(\gamma_k))_{\Omega} = \int_{\Omega_2} \psi_j \beta_l \int_{\Omega_1} \tau \phi_i \mathbf{c_x} \cdot \nabla_x \gamma_k dxdy.
\end{equation}
Then on the right we get:
\begin{equation}
    (\phi_i \psi_j, \tau \mathbf{c}_y \cdot \gamma_k \nabla_y(\beta_l))_{\Omega} = \int_{\Omega_2} \psi_j \nabla_y \beta_l \cdot \int_{\Omega_1} \tau \mathbf{c}_y \phi_i \gamma_k dxdy. 
\end{equation}

\printcredits


\bibliographystyle{elsarticle-num}

\bibliography{cas-refs}





\end{document}